\documentclass[11pt,a4paper,english]{article}
\usepackage{babel}
\usepackage{times}
\usepackage{amssymb}
\usepackage{epsfig}
\usepackage{natbib}
\bibpunct{[}{]}{;}{a}{,}{,}

\begin{document}
\title{Evaluation of Mutual Information Estimators for Time Series}

\author{Angeliki Papana\footnote{Email: agpapana@gen.auth.gr} , Dimitris Kugiumtzis \\
Department of Mathematical, Physical and Computational Sciences \\
Faculty of Engineering, Aristotle University of Thessaloniki \\
54124 Thessaloniki, Greece}
\maketitle
\bibliographystyle{apa}

\begin{abstract}
We study some of the most commonly used mutual information
estimators, based on histograms of fixed or adaptive bin size,
$k$-nearest neighbors and kernels, and focus on optimal selection
of their free parameters. We examine the consistency of the
estimators (convergence to a stable value with the increase of
time series length) and the degree of deviation among the
estimators. The optimization of parameters is assessed by
quantifying the deviation of the estimated mutual information from
its true or asymptotic value as a function of the free parameter.
Moreover, some common-used criteria for parameter selection are
evaluated for each estimator. The comparative study is based on
Monte Carlo simulations on time series from several linear and
nonlinear systems of different lengths and noise levels. The
results show that the $k$-nearest neighbor is the most stable and
less affected by the method-specific parameter. A data adaptive
criterion for optimal binning is suggested for linear systems but
it is found to be rather conservative for nonlinear systems. It
turns out that the binning and kernel estimators give the least
deviation in identifying the lag of the first minimum of mutual
information from nonlinear systems, and are stable in the presence
of noise.
\end{abstract}

{\em running title:} Mutual Information Estimators

\newpage

\section{Introduction}
Mutual information (MI) is a nonlinear measure used in many
aspects of time series analysis, best known as a criterion to
select the appropriate delay for state space reconstruction
\citep{Kantz97}. It is also used to discriminate different regimes
of nonlinear systems \citep{Hively00,Naa02,Wicks07} and to detect
phase synchronization \citep{Schmid04,Kreuz07}. Besides nonlinear
dynamics, it is used in various statistical settings, mainly as a
distance or correlation measure in data mining, e.g. in
independent component analysis and feature-based clustering
\citep{Tourassi01,Priness07}.

Any estimate of MI, either between two variables or as a function
of delay for time series, is (almost always) positively biased
\citep{Treves95,Moddemeijer89,Paninski03,Micheas06}. For
numerical-valued variables, MI increases with finer partition
depending on the underlying distribution and the sample size.
Beyond the classical domain partitioning, other schemes have been
used to estimate the densities inherent in the measure of mutual
information, e.g. kernels, B-splines and $k$-nearest neighbors
\citep{Moon95,Diks02,Daub04,Kraskov04}.

There are generally few analytic results on MI. Expressions of MI
in terms of the correlation coefficient are obtained for some
known distributions, e.g. Gaussian and Gamma distribution
\citep{Pardo95,Hutter05}. Some statistical results on the mean,
variance and bias of the MI estimator using fixed partitioning can
be found in \citep{Roulston97,Abarbanel01}, but the distribution
of any MI estimator is not known in general. For chaotic systems
in particular, the discontinuity of the density function of their
variables does not allow for an analytic derivation of the
statistics of MI estimators. Therefore, comparisons of MI
estimators relies on simulation studies. In some studies the
estimators are tested in identifying correctly the lag of the
first minimum of MI \citep{Moon95,Cellucci05}. Another performance
criterion is the bias of estimators in the case of Gaussian
processes, where the true MI is known
\citep{Cellucci05,Trappenberg06}.

MI estimation involves one and two dimensional density estimation.
Density estimation has been studied extensively and different
methods have been suggested and compared in the statistical
literature, e.g. see \citep{Scott79,Freedman81,Silverman86}, but
it is still to be investigated whether these methods and the
suggested criteria for the selection of method specific parameters
are also suitable for MI estimation. This is the main objective of
this study. MI estimators have also been compared to other linear
or nonlinear correlation measures \citep{Palus95,Steuer02,Daub04},
but we do not pursue this here as direct comparison is not
possible due to the different scaling of the measures, even after
normalization. There are some comparative studies on the MI
estimators and the selection of their parameters, as well as on
their performance on both linear and nonlinear dynamical systems
\citep{Wand93,Steuer02,Nicolaou05,Khan07}. Here, we extend these
studies, we evaluate some of the most commonly used MI estimators,
examine their consistency and optimize the selection of their
parameters including criteria for parameter selection suggested in
the literature. As the estimation depends on the underlying time
series we use Monte-Carlo simulations on systems of white noise of
different distributions, stochastic linear systems and dynamical
non-linear systems (maps and flows of varying complexity). The
performance of each estimator is examined with respect to the time
series length, the distribution of noise and the noise level in
the systems.

We study here the performance of three types of estimators, i.e.
estimators based on histograms (with fixed or adaptive bin size),
$k$-nearest neighbors and kernels. All estimators vary in the
estimation of the densities at local regions and we investigate
the optimal parameter for the determination of the two-dimensional
partitioning. Based on the simulation results, we propose optimal
parameters for each MI estimator with regard to the complexity of
the system, the observational noise level and the time series
length.

The structure of the paper is as follows. In
Sec.~\ref{sec:Estimators}, we briefly discuss the estimators
considered in this study and in Sec.~\ref{sec:SetUp} we present
the evaluation procedure and the simulated systems. In
Sec.~\ref{sec:Results}, we give quantitative results on the
dependence of the estimators on the parameters, time series length
and noise, we propose optimal parameter selection and compare the
different MI estimators. Finally, in Sec.~\ref{sec:Discussion} we
discuss the results and draw conclusions.

\section{Mutual Information Estimators}
\label{sec:Estimators}
MI is a measure of mutual dependence between two random variables
and quantifies the amount of uncertainty about one variable
reduced when knowing the other. The MI of two continuous random
variables $X, Y$ has the form
\begin{equation}
{\cal I}(X,Y)=\int_X \int_Y
f_{X,Y}(x,y)\log_a{\frac{f_{X,Y}(x,y)}{f_X(x)f_Y(y)}}
\mbox{d}x\mbox{d}y, \label{eq:def1}
\end{equation}
where $f_{X,Y}(x,y)$ is the joint probability density function
(pdf) of $X$ and $Y$, whereas $f_X(x)$ and $f_Y(y)$ are the
marginal pdfs of $X$ and $Y$, respectively. The units of
information of ${\cal I}(X,Y)$ depend on the base $a$ of the
logarithm, e.g. bits for the base $2$ logarithm and nats for the
natural logarithm.

Assuming a partition of the domain of $X$ and $Y$, the double
integral in Eq.(\ref{eq:def1}) becomes a sum over the cells of the
two-dimensional partition
\begin{equation}
{\cal I}(X,Y) = \sum_{i,j}p_{X,Y}(i,j) \log_a
{\frac{p_{X,Y}(i,j)}{p_X(i)p_Y(j)}}, \label{eq:def2}
\end{equation}
where $p_X(i)$, $p_Y(j)$, and $p_{X,Y}(i,j)$ are the marginal and
joint probability mass functions over the elements of the one and
two-dimensional partition. In the limit of fine partitioning the
expression in Eq.(\ref{eq:def2}) converges to Eq.(\ref{eq:def1}).
This may partly justify the abuse of notation of MI for the
continuous and the discretized variables. It is always ${\cal
I}(X,Y) \ge 0$, with equality holding for independent variables,
and ${\cal I}(X,Y) \le H(X) \le \log_a{n}$ (Jensen inequality),
where $H(X)= \sum_{i=1}^n p(x_i) \log_a p(x_i)$ is the Shannon
entropy of $X$. For a time series $\{X_t\}_{t=1}^n$, sampled at
fixed times $\tau_s$, MI is defined as a function of the delay
$\tau$ assuming the two variables $X=X_t$ and $Y=X_{t-\tau}$, i.e.
${\cal I}(\tau)={\cal I}(X_t,X_{t-\tau})$.

The Shannon entropy is always misestimated due to finite sample
effects \citep{Grassberger88,Kantz96}, but we do not discuss MI in
terms of entropies here as the estimation of MI boils down to the
estimation of the densities in Eq.(\ref{eq:def1}) or probabilities
in Eq.(\ref{eq:def2}). The estimators of MI, denoted $I(\tau)$,
differ in the estimation of the marginal and joint probabilities
or densities, using binning \citep{Fraser86,Darbellay99}, kernels
\citep{Silverman86,Moon95} or correlation integrals
\citep{Diks02}, $k$-nearest neighbors
\citep{Paninski03,Kraskov04}, $B$-splines \citep{Daub04} or the
Gram-Charlier polynomial expansion \citep{Blinnikov98}. All these
estimators depend on at least one parameter. We present bellow the
three first estimators that are most widely used.

\subsection{Binning estimators}
The most common MI estimator is the naive equidistant binning
estimator (ED) that regards the partition of the domain of each
variable into a finite number $b$ of discrete bins (equidistant
partitioning). The probability at each cell or bin is estimated by
the corresponding relative frequency of occurrence of the samples
in the cell or bin. The number of bins for each variable is the
same, so that the parameter to be optimized is the number of bins
for the partition or equivalently the bin width. The computation
of this MI estimator is straightforward as it is directly
estimated from the one and two dimensional histograms.

A second binning estimator is the equiprobable binning estimator
(EP), which is derived by partitioning the domain of each variable
in $b$ bins of the same occupancy (equiprobable partitioning) but
different width. The equiprobable partitioning actually transforms
the sample univariate distribution to discrete uniform with $b$
components minimizing the effect of the univariate distribution on
the estimation.

\citet{Fraser86} suggested an estimator using an adaptive
partitioning. This method constructs a locally adaptive partition
of the two-dimensional plane. It starts with a partition of
equiprobable bins for each variable and makes finer partition in
areas where the joint probability density is non-uniform until the
joint distribution on the cells is approximately uniform. The
final partition is finer in dense regions whereas less occupied
regions are covered with larger cells. It was found in
\citep{Palus93,Cellucci05} that this complex algorithm does not
substantially improve the binning estimator and requires large
data sets to gain accuracy; therefore it is not included in the
current evaluation.

A different estimator making use of adaptive partitioning (AD) is
proposed by \citet{Darbellay99}. The partition consists of
rectangles specified by marginal empirical quantiles, which are
not uniform in the sense that they are not made of a grid of
vertical and horizontal lines, irrespectively of whether these
lines are equally spaced or not. The AD estimator builds such a
partition in a way that it achieves conditional independence on
the rectangles of the partition. The advantage of this estimator
is that it is data-adaptive and does not a priori determine the
number of bins in the partition. The AD estimator has a direct
dependence on $n$, which determines the roughness of the
partitioning in a somehow automatic way. In the abundance of data,
the AD estimator reaches a very fine partition that satisfies the
independence condition in each cell, so that the total number of
cells is very large and analogous to a fixed-partition with a
respectively large $b$. Note that the dependence of AD on $n$ is
not comparable to that of the fixed-bin estimators because it
involves a change of partitioning with $n$.

For any binning scheme, the MI estimator $I(\tau)$ is given by
Eq.(\ref{eq:def2}) where the variables are
$(X,Y)=(X_t,X_{t-\tau})$, the sum is referred to the partition of
the two-dimensional domain of $(X_t,X_{t-\tau})$ and $p_{X_t}$,
$p_{X_{t-\tau}}$, $p_{{X_t},X_{t-\tau}}$ are the marginal and
joint probability distributions defined for each cell of the
partition.

\subsection{$k$-nearest neighbor estimator}
\citet{Kraskov04} proposed an MI estimator (KNN) that uses the
distances of $k$-nearest neighbors to estimate the joint and
marginal densities. For each reference point from the bivariate
sample, a distance length is determined so that the $k$ nearest
neighbors are within this distance length. Then the number of
points within this distance from the reference point gives the
estimate of the joint density at this point and the respective
neighbors in one-dimension give the estimate of the marginal
density for each variable. The algorithm uses discs (or squares
depending on the metric) of a size adapted locally and then uses
the corresponding size in the marginal subspaces, so in some sense
the estimator is data adaptive. Still, it involves as a free
parameter the number of neighbors $k$. Note that a large $k$
regards a small $b$ of the fixed binning estimators. However, the
estimator does not use a fixed neighborhood size and therefore
there is not a clear association of $k$ and $b$. The KNN estimator
is data efficient, adaptive, has minimal bias and is recommended
for high-dimensional data sets \citep{Kraskov04}. It requires an
additional computational cost for the search of the $k$ neighbors.

\subsection{Kernel density estimator}
The kernel density MI estimator (KE) uses a smooth estimate of the
unknown probability density by centering kernel functions at the
data samples; kernels are used to obtain the weighted distances
\citep{Silverman86,Moon95}. The kernels essentially weigh the
distance of each point in the sample to the reference point
depending on the form of the kernel function and according to a
given bandwidth $h$, so that a small $h$ produces details in the
density estimate but may loose in accuracy depending on the data
size.

KE estimator has two free parameters, the bandwidth $h_1$ for the
marginal densities of $X$ and $Y$, and the bandwidth $h_2$ for the
joint density of $(X,Y)$. The bandwidth $h_1$ is related to $b$ of
the fixed binning estimators by an inverse relation, e.g. a
rectangular kernel assigns a bin centered at the reference point.
Its advantage over binning estimators is that the location of the
bins is not fixed. Among the different kernel functions, Gaussian
kernels are most commonly used and we use them here as well. A
kernel density estimator with Gaussian kernel function and a fixed
bandwidth $h$ at a point $\textbf{x} \in R^d$ is
\begin{equation}
f(\textbf{x})=\frac{1}{{{(2\pi)}^{d/2}h^d \mbox{det}(\textbf{S})^{1/2}}}
\sum_{i=1}^{n^\prime} \exp\left({-\frac{{(\textbf{x}-\textbf{x}_i)}^\tau
\textbf{S}^{-1}(\textbf{x}-\textbf{x}_i)}{2h^2}}\right),
\label{eq:densitykernel}
\end{equation}
where $\textbf{S}$ is the data covariance matrix and $n^\prime$ is
the number of the $d$-dimensional vectors \citep{Moon95}.

\section{Simulation Setup}
\label{sec:SetUp}
The evaluation of the estimators is assessed by Monte-Carlo
simulations on white noise, linear systems and chaotic systems of
different complexity, listed in Table~\ref{tab:systems}. \\
=============================================== \\
  TABLE 1      ***  To be placed here   ***     \\
=============================================== \\
A Gaussian and a skewed Gamma distribution are used to generate
white noise time series, whereas for linear systems, the
autoregressive model AR(1) and autoregressive moving average
ARMA(1,1) are used with coefficients as given in
Table~\ref{tab:systems}, assuming both Gaussian and Gamma input
white noise. The non-linear chaotic systems are the Henon map
\citep{Henon76}
\[x_t = 1-ax_{t-1}^2+bx_{t-2},\]
the Ikeda map \citep{Ikeda80}
\begin{eqnarray*}
x_t & = & a + b (x_{t-1} \cos u_{t-1} - y_{t-1} \sin u_{t-1}) \\
y_t & = & b (x_{t-1} \sin u_{t-1} + y_{t-1} \cos u_{t-1}),
\end{eqnarray*}
where $u_t = \kappa - \frac{\eta}{1+x_t^2+y_t^2}$, and the
Mackey-Glass differential system \citep{Mackey77}
\[\frac{\mbox{d}x}{\mbox{d}t} =
\frac{0.2x_{t-\Delta}}{1+x_{t-\Delta}^{10}}-0.1x_t\]
where the delay $\Delta$ accounts for the system complexity. We
use here $\Delta=17$ and $\Delta=30$ for low-dimensional chaos of
fractal dimension about 2 and 3, respectively, and $\Delta=100$
for high-dimensional chaos of fractal dimension about 7.
Observational white noise at different levels is also assumed for
the chaotic systems, given as a percentage of the standard
deviation of the noise-free data.

Different lengths $n$ for the generated time series from each
system are considered as follows. For white noise and linear
systems, $n$ is given in powers of $2$ from $5$ to $13$ and for
nonlinear systems from $8$ to $13$. $I(\tau)$ is computed using
all methods on $1000$ realizations for each system, noise type or
level, and time series length. As all linear systems are of order
$1$, MI is computed only for lag $1$. For the nonlinear systems,
$I(\tau)$ is computed up to the lag $\tau$ for which $I(\tau)$
levels-off. For the Mackey-Glass system we compute the lag of the
first minimum of $I(\tau)$ and specifically for $\Delta=100$ the
lag that MI levels-off because it does not exhibit a distinct
minimum. For each estimator, $I(\tau)$ is computed for a wide
range of values of the free parameter and for specific values
determined by standard criteria, which are specified below.

For the binning estimators ED and EP we set the number of bins to
$b=2,4,8,16,32,64$. We also consider $10$ commonly used criteria
for $b$, given in Table~\ref{tab:binnumber}.    \\
=============================================== \\
  TABLE 2      ***  To be placed here   ***     \\
=============================================== \\
For the choice of $k$ of $k$-nearest neighbor estimator KNN,
\citet{Kraskov04} propose to use $k=2$ to $4$ (these are also used
in \citep{Kreuz07,Khan07}). However for real world data one should
investigate also larger values of $k$. Therefore we use in the
simulations a wide range of $k$ values as for $b$,
$k=2,4,8,16,32,64$.

Among different kernel functions used in the literature for
density estimation, and for MI estimation in particular, the
common practice is to use the Gaussian kernel in conjunction with
the "Gaussian" bandwidth of \citet{Silverman86}
\begin{equation}
\ h = {\left(\frac{4}{(d+2)n^\prime}\right)}^{1/(d+4)}
\label{eq:gaussianbandwidth}
\end{equation}
($n^\prime$ is the number of the $d$-dimensional vectors) or
multiples of it \citep{Harrold01,Steuer02,Khan07}. In the
estimation of mutual information with kernels, the range of
bandwidths is usually not searched and a bandwidth is selected
according to a criterion such as the "Gaussian" bandwidth
\citep{Moon95,Steuer02}. A multiple bandwidth selection scheme for
the test for independence is proposed in \citep{Diks08}. Analytic
and simulation studies have shown that the choice of the bandwidth
is crucial and depends on the data size
\citep{Bonnlander94,Jones96}. Therefore we consider a wide range
for the bandwidth $h_1$ for one dimension and $h_2$ for two
dimensions, as for $b$ and $k$. Specifically, for $h_1$ we take
$15$ values in $[0.01,2]$ at a fixed base-2 logarithmic step and
set $h_2=h_1$ and $h_2=\sqrt{2}h_1$. The second form for $h_2$
accounts for the scaling of the Euclidean metric in $\Re^2$, which
we use in the simulations. We also consider some well-known
criteria for the choice of bandwidths, given in
Table~\ref{tab:bandwidth}.                      \\
=============================================== \\
  TABLE 3      ***  To be placed here   ***     \\
=============================================== \\
The three first criteria define bandwidth for both one and
two-dimensional space. For the other criteria we set $h_2$ equal
either to $h_1$ or $\sqrt{2}h_1$.

The true (theoretical) MI ${\cal I}(\tau)$ is not known in
general. However, for Gaussian processes ${\cal I}(\tau)$ is given
in terms of the autocorrelation function $\rho(\tau)$ as
\begin{equation}
{\cal I}(\tau)= -0.5\log(1-{\rho^2(\tau)}).
 \label{eq:trueMI}
\end{equation}
In the lack of the true MI for the other systems, we assume
consistency of the estimators and use the asymptotic value of
$I(\tau)$ computed on a realization of size $n=10^7$. In this
computation, we set for the ED and EP estimators $b=64$ (we also
computed MI for $b$ up to 256, however MI did not substantially
differed), for KNN estimator $k=2$, and for the KE estimator
$h_1=0.01$ and $h_2=h_1$. Similar approach for approximating the
true MI is used in \citep{Harrold01,Cellucci05}. We denote
$I_{\infty}$ the true or asymptotic MI and use it as reference to
compute the accuracy of the different estimators.

We evaluate the estimators and their parameters separately for
white noise and linear systems, for the nonlinear maps Henon and
Ikeda, and for the Mackey-Glass system. First, we investigate the
dependence of the estimators on their free parameter for white
noise and linear systems and for different time series lengths
giving a total of $L=126$ cases (14 systems and 9 time series
lengths). For each estimator, we compute the mean estimated MI
$\bar{I}_c(l)$ from the $1000$ realizations for each tested value
of the free parameter denoted by $c$, where l denotes the system
case and $l=1,...,L$. Further, for each $l$ we compute the
deviation $dI_c(l)=|\bar{I}_c(l)-I_{\infty}(l)|$, where
$I_{\infty}(l)$ is either the true MI (for Gaussian processes) or
the asymptotic value computed from each estimator. All estimators
converged to the same MI under proper parameters as obtained for
$n$ increasing up to $10^7$. Given the asymptotic MI
$I_{\infty}(l)$, the optimal parameter values for each case $l$
(system and time series length) is obtained from the minimum
deviation $dI_c(l)$ with respect to $c$. The estimators are then
compared for their optimal parameters by computing the divergence
of the mean $I(\tau)$ of each estimator from $I_{\infty}(l)$ for
all cases.

For the discrete nonlinear systems, there are $L=48$ cases ($8$
maps including different noise levels and $6$ time series
lengths). Here, a single $I_{\infty}$ for each system cannot be
obtained as the MI estimate for $n$ increasing up to $10^7$ does
not converge and the MI for $n=10^7$ is still dependent on
parameter selection and varies also across estimators. Therefore,
for each system we set as asymptotic value $I_{\infty}$ of the
estimator the MI computed for $n=10^7$ and for a very fine
partition. So here, the interest is in the dependence of each
estimator on the free parameter and the rate of convergence
towards the asymptotic value.

For the nonlinear flows derived from the Mackey-Glass system for
$\Delta=17,30$ (noise-free and with noise), we concentrate on the
first minimum of MI and compute the lag $\tau_0$ of the first
minimum of MI for each of the $1000$ realizations of each system.
For $\Delta =100$ there is no clear minimum of MI and it follows a
rather exponential decay. Therefore we compute instead the lag
$\tau_0$ for which MI levels off according to a criterion for
levelling. In order to compare the estimators, we examine the
consistency of each estimator with $n$, the dependence of the
estimation of $\tau_0$ on the parameter selection and the variance
of the estimated lags $\tau_0$ from all cases.

\section{Results}
\label{sec:Results}
\subsection{Results on white noise and linear systems}
\begin{figure}[h!]
\centerline{\hbox{
\includegraphics[height=9cm,keepaspectratio]{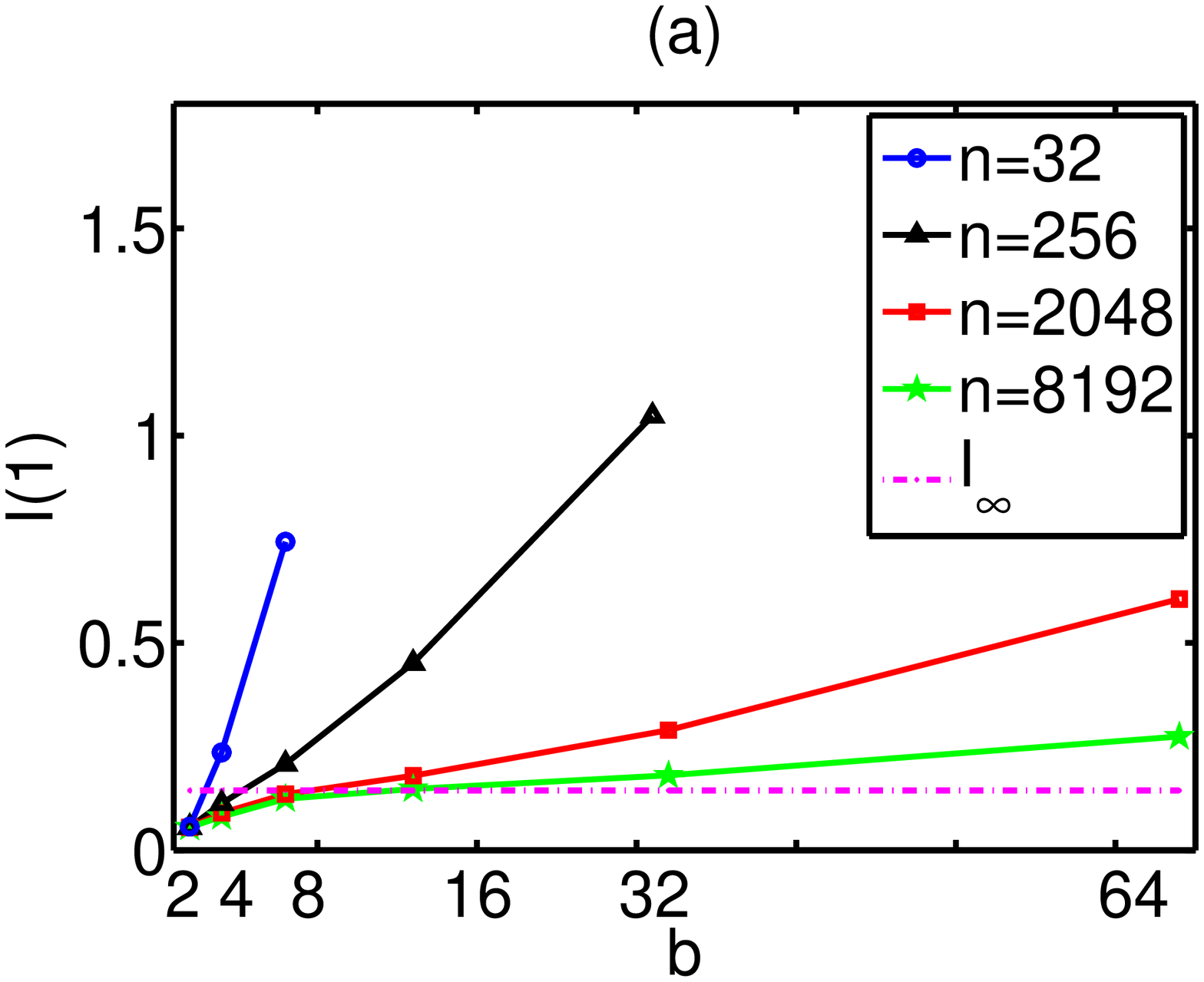}}}
\centerline{\hbox{
\includegraphics[height=9cm,keepaspectratio]{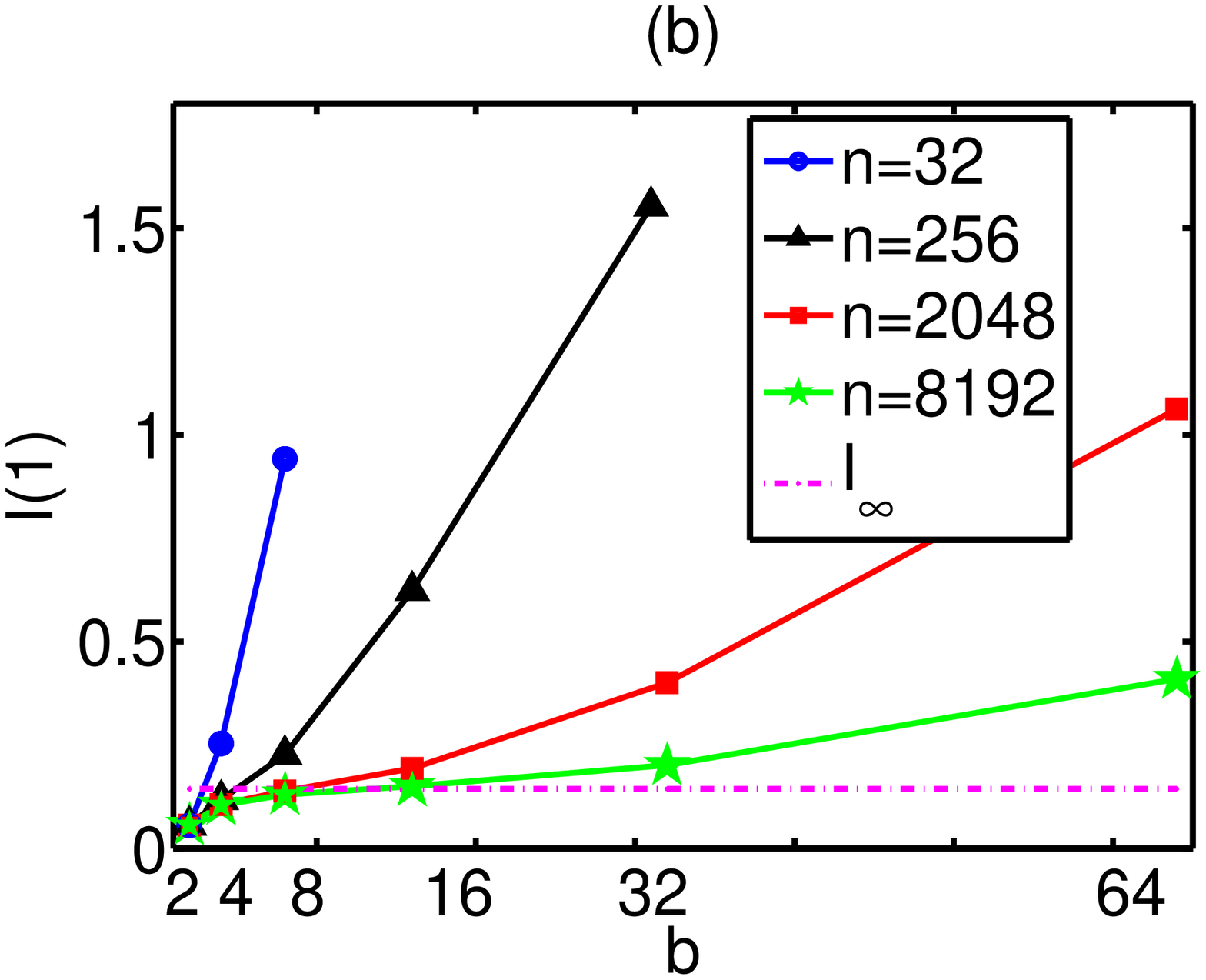}}}
\caption{A. Papana} \label{fig:AR1r05MI}
\end{figure}

The MI for lag one $I(1)$ from the binning estimators ED and EP
increases always with the number of bins $b$. Thus for white noise
where $I_{\infty}(1)=0$ the best choice for $b$ is 2 that gives
the smallest positive $I(1)$. For the linear processes, $I(1)$
decreases with the time series length $n$ for each $b$, as shown
in Fig.~\ref{fig:AR1r05MI} for the ED and EP estimates of $I(1)$
from an AR(1) process.                          \\
=============================================== \\
  Figure 1       ***  To be placed here   ***   \\
=============================================== \\
We note that for sufficiently large $b$, $I(1)$ converges with $n$
to the true value ${\cal I}(1)=I_{\infty}(1)$ given in
(\ref{eq:trueMI}). On the other hand, for small $b$, $I(1)$
underestimates $I_{\infty}(1)$ depending again on $n$. Thus the
optimal $b$ that gives the smallest $dI_c(l)$ and estimates best
$I_{\infty}(1)$ depends on $n$.

\begin{figure}
\centerline{\hbox{
\includegraphics[height=9cm,keepaspectratio]{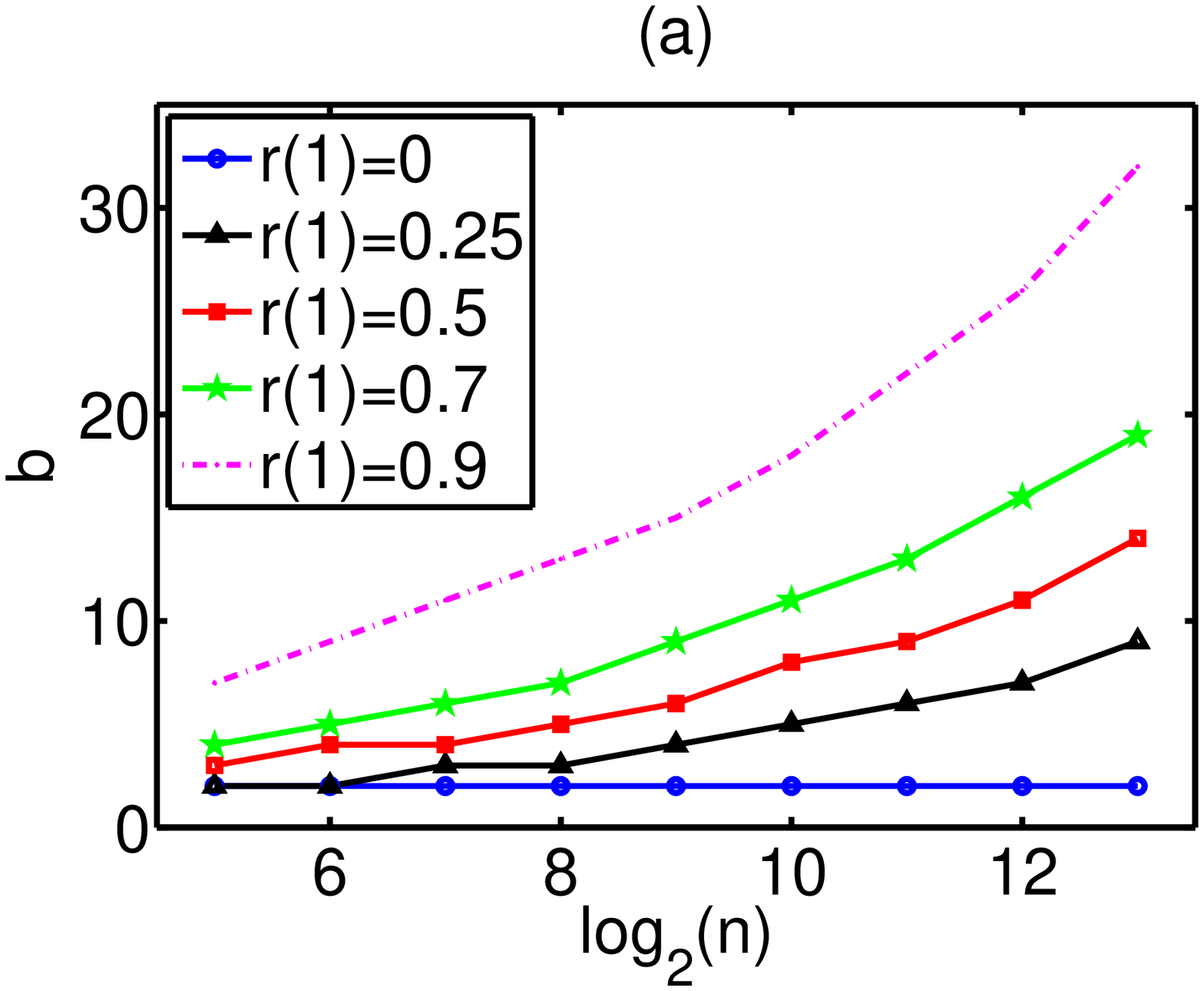}}}
\centerline{\hbox{
\includegraphics[height=9cm,keepaspectratio]{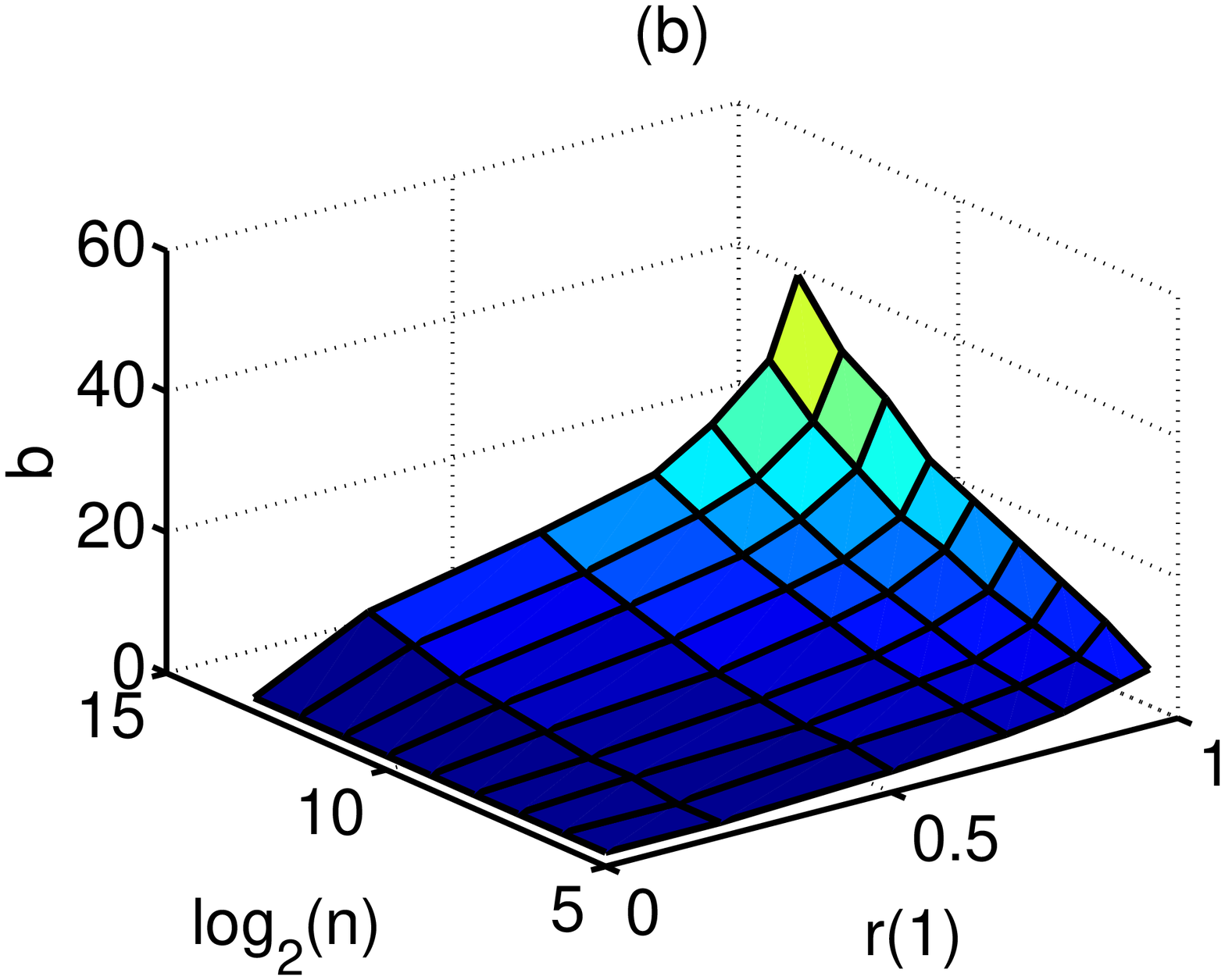}}}
\caption{A. Papana} \label{fig:linearoptbin}
\end{figure}

We have observed that $b$ depends also on the autocorrelation
function $r(\tau)$ of the linear system. To investigate further
this dependence we computed ED and EP estimates of $I(1)$ for a
wide range of $r(1)$ values of an AR(1) process. We found that the
optimal $b$ (found by the smallest $dI_c(l)$) increases smoothly
with $n$ and $r(1)$, as shown in Fig.~\ref{fig:linearoptbin}.  \\
=============================================== \\
  Figure 2       ***  To be placed here   ***   \\
=============================================== \\
A search for a parametric fit of optimal $b$ regarding the graph
of Fig.~\ref{fig:linearoptbin}b resulted in the form
\begin{equation}
b = \alpha n^{\beta} e^{\gamma \rho^2}
 \label{eq:numbins}
\end{equation}\
where the coefficients $\alpha$, $\beta$, $\gamma$ take similar
values for the ED and EP estimators (0.65,0.25,2.11 and
0.76,0.19,1.91 respectively).

Most of the criteria in Table~\ref{tab:binnumber} tend to
overestimate $b$. To evaluate the performance of the 10 criteria
in Table~\ref{tab:binnumber} and the proposed criterion in
Eq.(\ref{eq:numbins}), we compute the total score $S_c$ of each
criterion $c$, where $c=1,\ldots,11$, for all $L=126$ tested
systems and time series lengths, as
\begin{equation}
S_c =  \frac {\sum_{l=1}^L (\bar{I}_c(l)-I_{\infty}(l))^2}
{\sum_{l=1}^L (\bar{\bar{I}}_c(l)-I_{\infty}(l))^2}
\label{eq:scores}
\end{equation}\
where for each case $l$, $\bar{\bar{I}}_c(l)= \frac{1}{11}
\sum_{c=1}^{11} \bar I_c(l)$ is the grand mean of the means from
all criteria. According to the score $S_c$, the proposed criterion
in Eq.(\ref{eq:numbins}) for the optimal $b$, denoted H11,
outperforms the other criteria when ED estimator is used, as shown
in Table~\ref{tab:optCriteria}.                 \\
=============================================== \\
  Table 4      ***  To be placed here   ***     \\
=============================================== \\
For the EP estimator, criterion H9 scores lowest and H11 is ranked
fifth but the differences in the scores of the best five criteria
are comparatively small.

\begin{figure}
\centerline{\hbox{
\includegraphics[height=9cm,keepaspectratio]{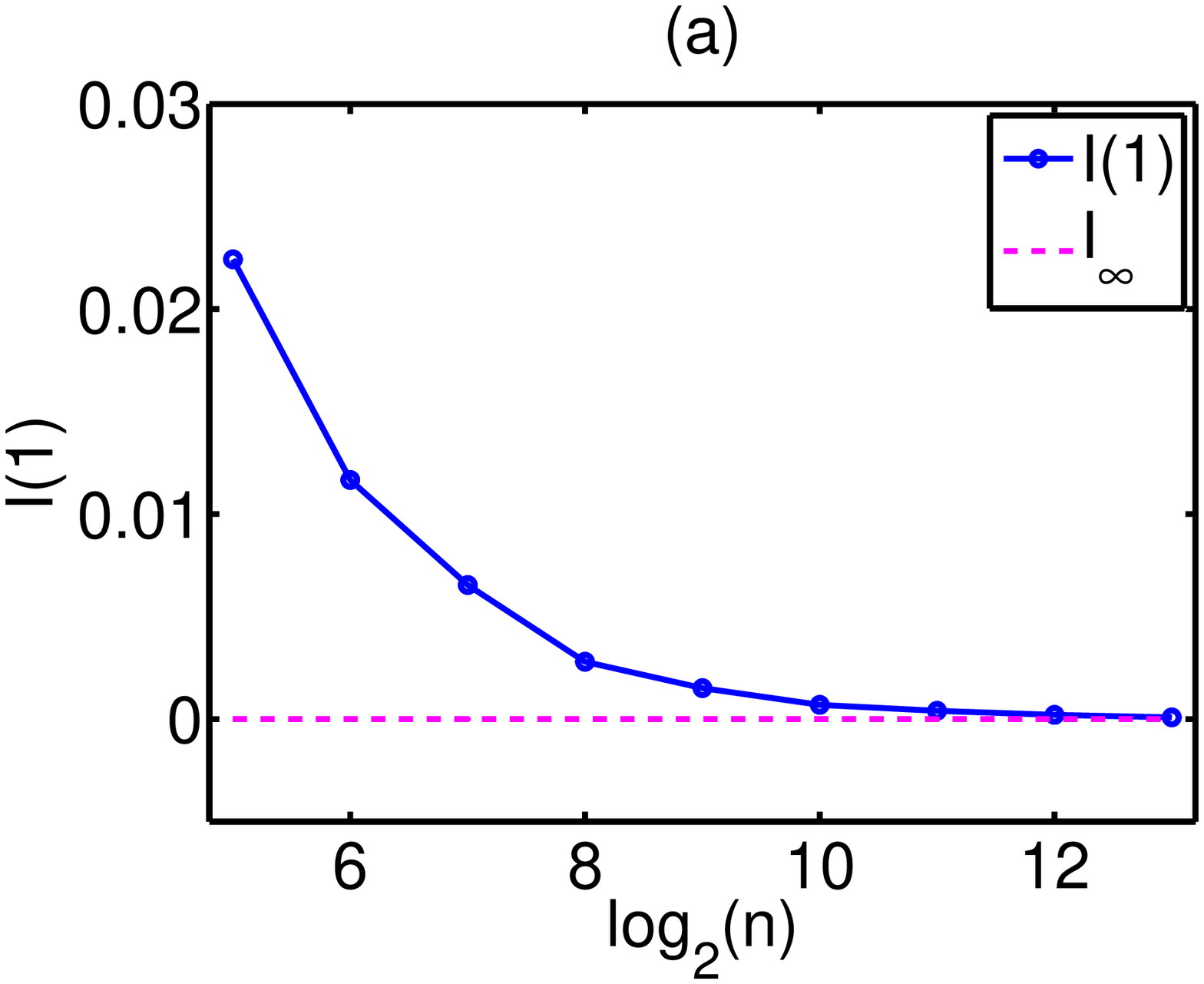}}}
\centerline{\hbox{
\includegraphics[height=9cm,keepaspectratio]{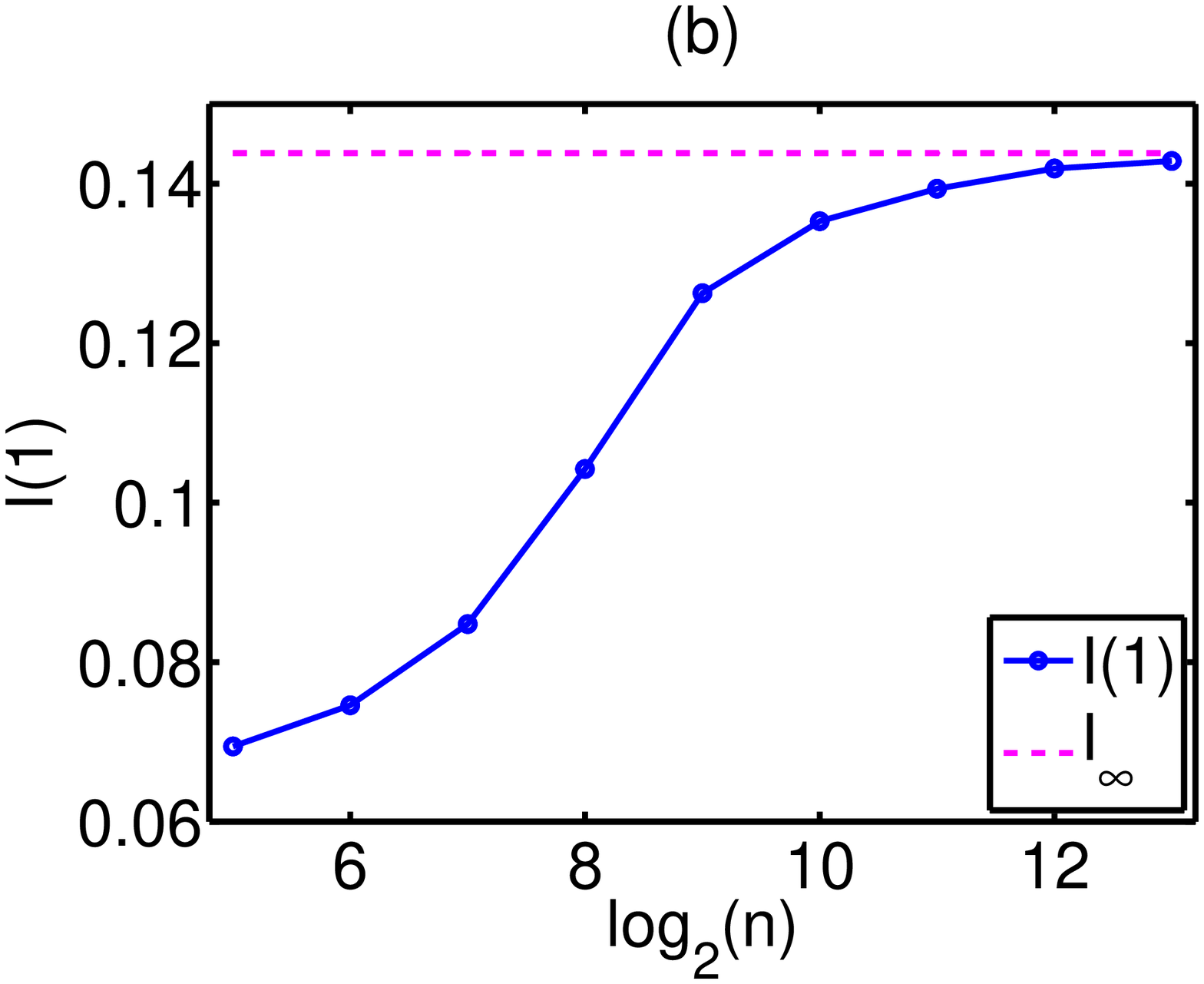}}}
\caption{A. Papana} \label{fig:MIADlinear}
\end{figure}

For certain bivariate distributions and Gaussian processes, it was
found that the AD estimator was precise in estimating MI and
converged fast to the true MI \citep{Kraskov04,Trappenberg06}. We
confirmed this result by our simulations on the white noise and
linear systems with the remark that the convergence to
$I_{\infty}$ is rather slow and is succeeded at large $n$, as
shown in Fig.~\ref{fig:MIADlinear}.
=============================================== \\
  Figure 3       ***  To be placed here   ***   \\
=============================================== \\

\begin{figure}
\centerline{\hbox{
\includegraphics[height=9cm,keepaspectratio]{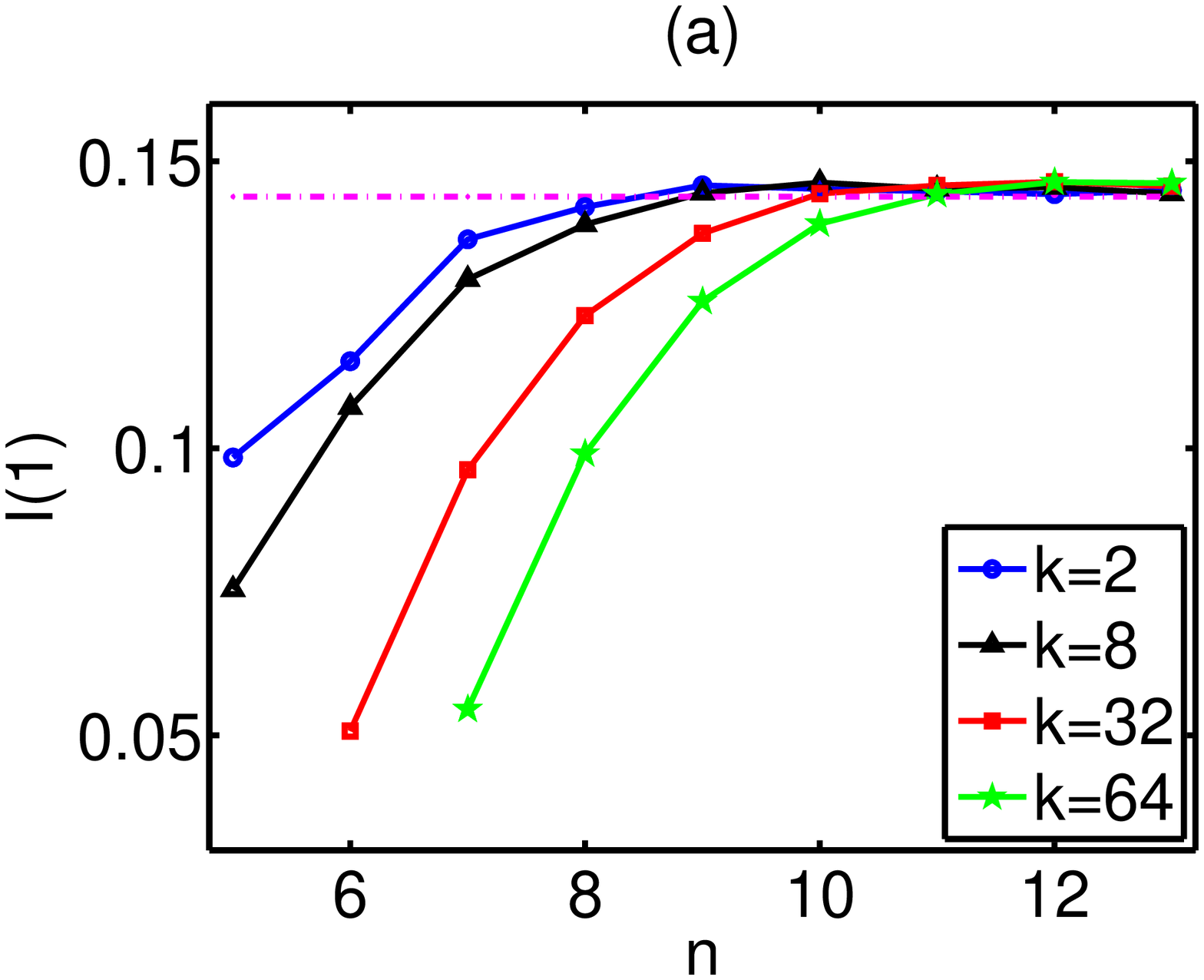}}}
\centerline{\hbox{
\includegraphics[height=9cm,keepaspectratio]{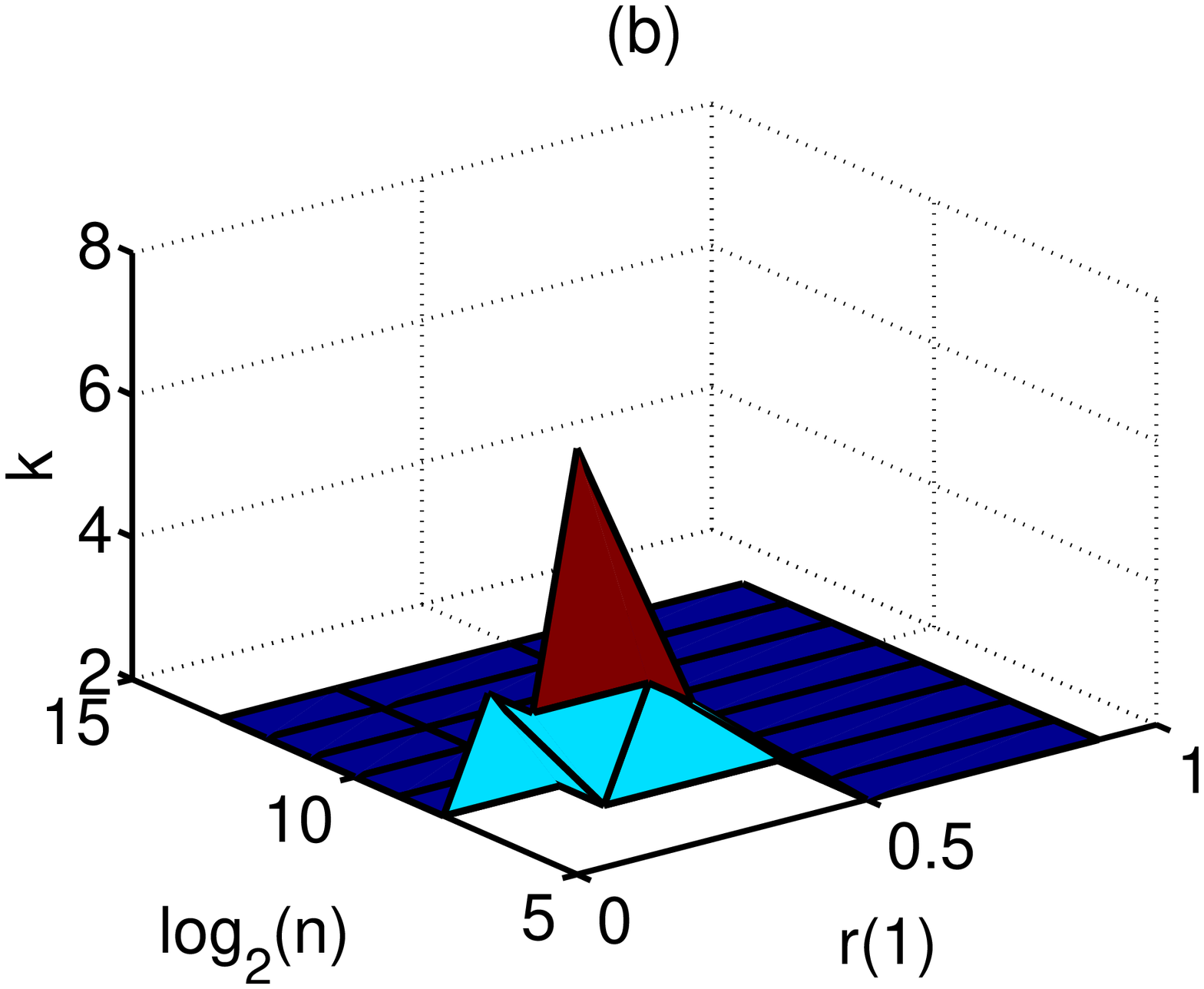}}}
\caption{A. Papana} \label{fig:KNNAR1r05norm}
\end{figure}

The number of nearest neighbors $k$ in the KNN estimator
determines the roughness of approximation of the density functions
in Eq.(\ref{eq:def1}), which corresponds to the roughness of the
partitioning in Eq.(\ref{eq:def2}). The simulations showed that
for white noise the MI estimated by KNN is close to zero for a
long range of $k$ and the deviation from zero decreases as $k$
approaches $n/2$ (as reported also in \citep{Kraskov04}). For the
linear systems the optimal $k$ is rather small. The dependence of
the KNN estimator on $k$ holds mainly for small time series as for
large $n$ the estimated MI converges to $I_{\infty}$ for any $k$,
as shown in Fig.~\ref{fig:KNNAR1r05norm}a.      \\
=============================================== \\
  Figure 4       ***  To be placed here   ***   \\
=============================================== \\
Still, the convergence is slower for larger $k$. In any case, a
highly accurate MI is attained with small $k$ for all but very
small time series. For example, for an accuracy threshold of
$10^{-4}$ in estimating $I_{\infty}$, i.e. $|I(1) -
I_{\infty}|<10^{-4}$, the optimal choice for $k$ is 2 in almost
all cases except for very small $n$ and $r(1)$, as shown in
Fig.~\ref{fig:KNNAR1r05norm}b. Note that even for white noise time
series of small length, $k \le 8$ reaches this accuracy threshold
(the peak in the graph of Fig.~\ref{fig:KNNAR1r05norm}b is for
$n=2^5$ and $r(1)=0$).

\begin{figure}
\centerline{\hbox{
\includegraphics[height=9cm,keepaspectratio]{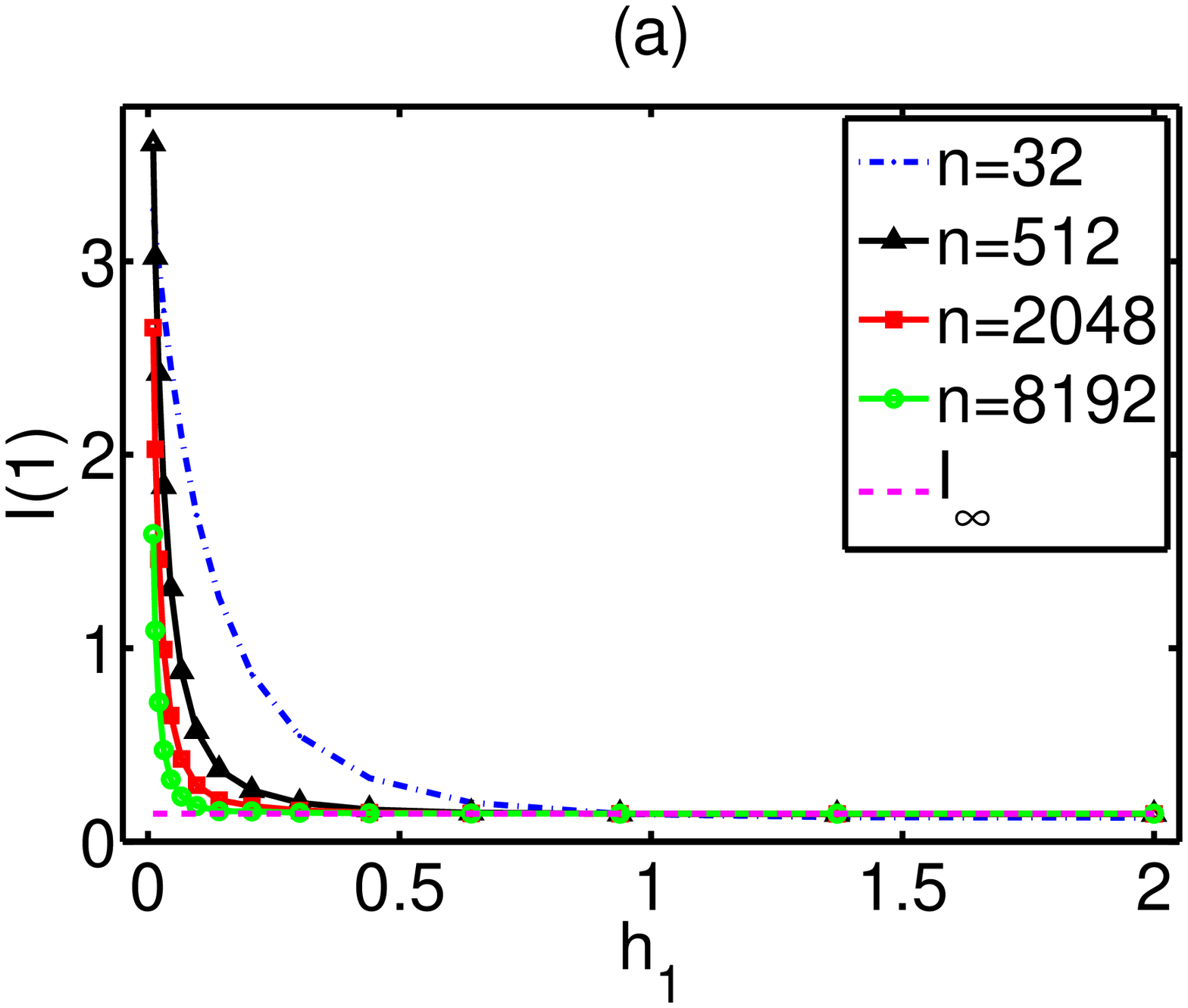}}}
\centerline{\hbox{
\includegraphics[height=9cm,keepaspectratio]{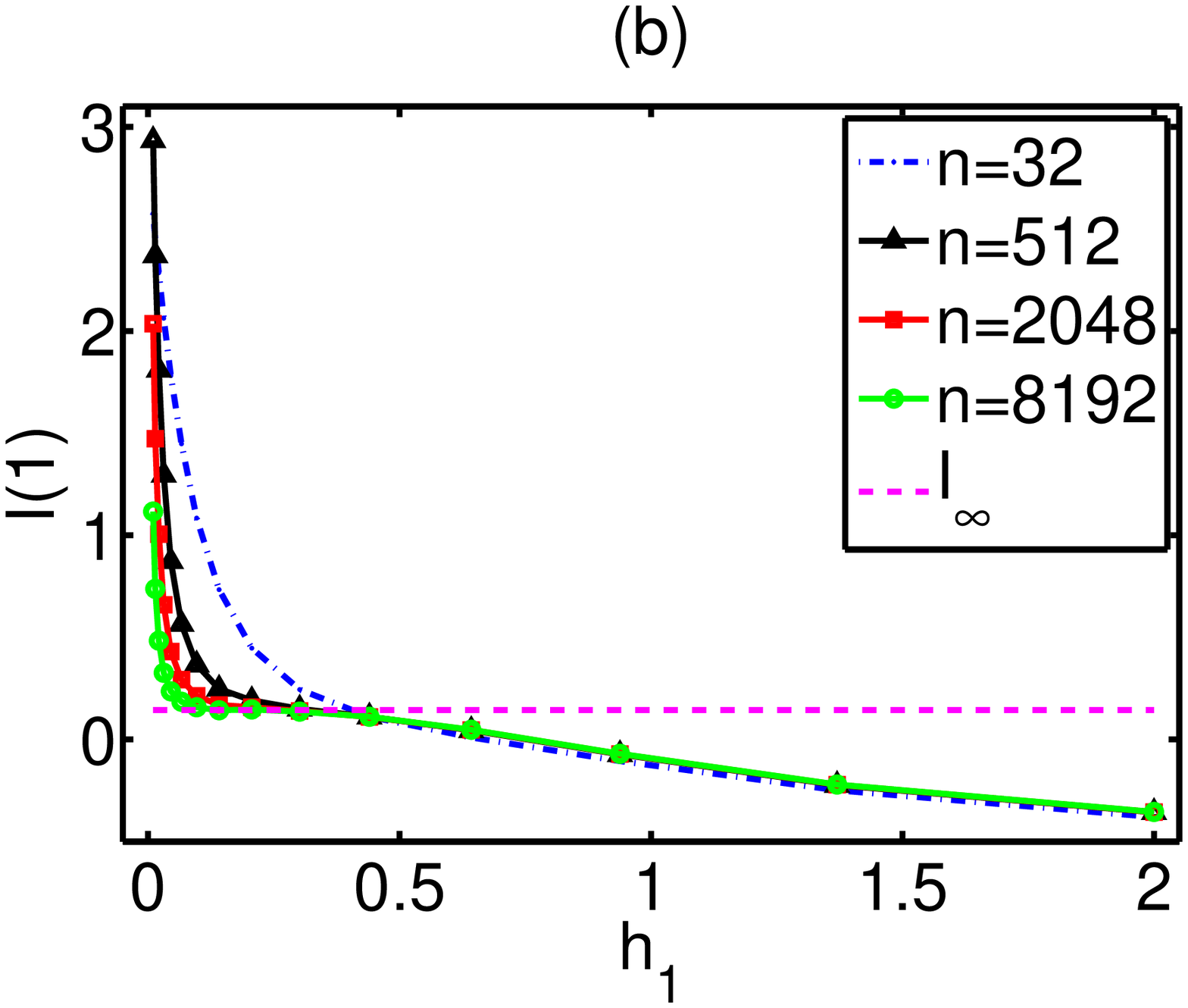}}}
\caption{A. Papana} \label{fig:MIKE}
\end{figure}

For the two dimensional bandwidth $h_2$ we have considered
$h_2=h_1$ and $h_2=\sqrt{2}h_1$ and studied the dependence of the
estimated MI on $h_1$ and $h_2$ across a large range of bandwidths
for white noise and linear systems. As for $k$ of the KNN
estimator, MI converges with $n$ and faster for smaller $h_1$. For
$h_2= h_1$ the convergence with $n$ is correctly towards
$I_{\infty}$ (see Fig.~\ref{fig:MIKE}a), but for $h_2=\sqrt{2}h_1$
MI decreases with $h_1$ and becomes negative (see
Fig.~\ref{fig:MIKE}b).                          \\
=============================================== \\
  Figure 5       ***  To be placed here   ***   \\
=============================================== \\

This result advocates the use of the same bandwidth for the kernel
estimates of the marginal and joint distributions. We have also
investigated whether there is dependence of $h_1$ on $r(1)$ and
$n$. As for $k$ of the KNN estimator, there does not seem to be
any systematic dependence. Using the same threshold accuracy in
estimating $I_{\infty}$, the smallest optimal $h_1$ is always at a
low level for all $r(1)$ and $n$ and there is no apparent pattern
that would suggest a particular form of dependence of $h_1$ on
$r(1)$ or $n$, as shown in Fig.~\ref{fig:MIKE}c. The sudden jumps
in the graph of Fig.~\ref{fig:MIKE}c is due to numerical
discrepancies around the chosen threshold for different $h_1$
values.

In order to evaluate the criteria for selecting $h_1$ and $h_2$ in
Table~\ref{tab:bandwidth}, we computed for each criterion the
score defined in Eq.(\ref{eq:scores}) for all cases. The five
optimal criteria and their scores for varying lengths of time
series from white noise and linear systems are C1 ($0.85$), C3
($0.94$), C2 ($0.96$), C4 ($1.57$) and C9 ($1.67$). The simplest
criteria turned out to score lowest with best being the "Gaussian"
rule of Silverman C1 (see also Eq.(\ref{eq:gaussianbandwidth})).

Summarizing the results on white noise and linear systems, it
turns out that fixed binning estimators are the most dependent on
the free parameter, the number of bins $b$, whereas for KNN and KE
estimators a small number of neighbors $k$ and bandwidth $h_1$,
respectively, turns out to be sufficient for all but very small
time series length $n$ and weak autocorrelation $r(\tau)$. In such
cases, binning estimators can approximate $I_{\infty}$ better with
a relatively small $b$ and we provided an expression for this
involving $n$ and $r(1)$. All estimators are consistent but
converge at different rates to the true or asymptotic MI
$I_{\infty}$, as shown in Fig.~\ref{fig:alllinear} for the AR(1)
system with weak and strong autocorrelation.    \\
=============================================== \\
  Figure 6       ***  To be placed here   ***   \\
=============================================== \\

In general, the KNN estimator converges fastest. To this respect,
the parameter-free AD estimator would be the second best choice
after the KNN estimator because it showed a slower convergence
rate. KE estimator has about the same convergence rate as AD and
is not significantly affected by parameter selection (for
$h_2=h_1$), however it would not be preferred due to its
computational cost. The estimation accuracy of each estimator is
quantified by the index $dI_{c_{opt}} = \sum_{i=1}^L
(\bar{I}_{c_{opt}}(l)-I_{\infty}(l))^2$, where $L=126$ and
$c_{opt}$ is the optimal free parameter found in the simulations
above, i.e. H11 for $b$ in ED and H9 for $b$ in EP, $k=2$ for KNN,
and C1 for the bandwidths in KE. The smallest index
$dI_{c_{opt}}=0.254$ was obtained by ED, followed by KE (0.302)
and KNN (0.496), whereas AD and EP scored worse (1.865 and 2.231,
respectively). Our numerical analysis on the linear systems and
noise showed that for the three aspects of estimation considered,
i.e. parameter dependence, rate of convergence, and accuracy of
estimation, no estimator ranks first but KNN and KE turn out to
perform overall best.

\subsection{Results on nonlinear maps}
In terms of chaotic systems, let us first note that MI can be
viewed as a measure on the reconstructed attractor projected on
$\Re^2$, i.e. on points $[x_t,x_{t-\tau}]'$. Due to the fractal
structure in all scales of the chaotic attractor, ${\cal I}(\tau)$
defined in terms of a partition (see Eq.(\ref{eq:def2})) increases
with finer partition towards the limit of ${\cal I}(\tau)$ given
in Eq.(\ref{eq:def1}) for the continuous space. On the other hand,
for the estimation of entropy, and particularly the
Kolmogorov-Sinai or metric entropy, it is postulated that there
exists a so-called generation partition that gives the expected
entropy value and further refinement to this partition does not
increase further the computed entropy \citep{Walter75, Cohen85}.
However, with regard to the Shannon entropy, we observed that we
can only get an upper limit of MI from the KNN estimator with
$k=2$ as the estimation algorithm does not allow for a finer
partition, whereas increasing $b$ for the binning estimators or
decreasing $h_1$ for the KE estimator within the tested range does
not seem to lead to convergence of MI.

The true ${\cal I}(\tau)$ in Eq.(\ref{eq:def1}) is not known since
the joint distribution of $[x_t,x_{t-\tau}]'$ is also not known.
This prevents the direct comparison of the estimators and the
search for optimal free parameters. The presence of noise in
chaotic time series sets a limit to the scale where fractal
details can be observed and consequently to the finiteness of the
partition when estimating ${\cal I}(\tau)$. In that case, an
asymptotic $I_{\infty}$ does exist and the performance of the
estimators in terms of the free parameter and time series length
can be compared, also for different noise levels. In the
following, we try to delineate the differences among estimators in
estimating $I_{\infty}$ and particularly in converging to
$I_{\infty}$ with respect to the time series length and their free
parameter.

The discussion above would suggest that the estimated MI should
always increase as the partition gets finer, but in practice this
requires a sufficient time series length $n$. For the binning
estimators the optimal number of bins $b$, i.e. the $b$ giving
largest MI and minimum $|I_{\infty}(\tau) - \bar{I}_c(\tau)|$, is
not always the largest (limited to $b=64$ in our study) but
increases with $n$, as shown in Fig.~\ref{fig:henonoptbin}a for
the ED estimator and the noise-free Henon map.  \\
=============================================== \\
  Figure 7       ***  To be placed here   ***   \\
=============================================== \\
In the same figure the limits for optimal $b$ from the suggested criterion
H11 in Eq.(\ref{eq:numbins}) (lower for r$(1)=0$ and upper for r$(1)=1$)
are shown with dotted lines and are well beyond the optimal bins found
for small lags. For this system, $I(\tau)$ decreases smoothly with $\tau$ and
therefore the optimal $b$ decreases as well. For $\tau=10$,
$I(\tau)$ levels off for small $n$ and then $b \simeq 2$ is
optimal, but as $n$ increases more bins give indeed larger values
of MI. Thus as $n$ increases weak MI for large $\tau$ becomes
significant and can be distinguished from the plateau of
independence only when a larger $b$ is used for the binning
estimator. However, for a fixed $b$ MI converges to $I_{\infty}$
with $n$, even for noise-free data (Fig.~\ref{fig:henonoptbin}c).

With the addition of noise, $I(\tau)$ decreases and the optimal
number of bins drops, as shown in Fig.~\ref{fig:henonoptbin}b and
d respectively for $20\%$ additive noise on the Henon time series.
The stronger the noise component is, the more the deterministic
structure is masked and the faster the estimated MI levels towards
zero with the lag. For the noisy chaotic data, the pattern of the
dependence of the binning estimates of MI to $n$ and $b$ is closer
to the one observed for the linear systems. For example, the range
of optimal $b$ in Fig.~\ref{fig:henonoptbin}b is at the level of
$b$ given by the suggested criterion H11 for $r(1)$ ranging from
0 to 1. The results on EP estimator are similar.

In line with the ED and EP estimators, the AD estimator does not
converge with $n$ to $I_{\infty}$ for the nonlinear systems unless
the fine partition is limited by the presence of noise, as shown
in Fig.~\ref{fig:MIAD} for the Henon map.       \\
=============================================== \\
  Figure 8       ***  To be placed here   ***   \\
=============================================== \\
The increase of $n$ directs the algorithm of AD to make a finer
partition which results in a larger $I(\tau)$. The effect of $n$
on the adaptive estimator decreases with the increase of the noise
level.

As pointed earlier, there is a loose relationship between the
number of nearest neighbors $k$ in the KNN estimator and the
number of bins $b$ in the binning estimators, i.e. small $k$
corresponds to large $b$. The lower limit $k=1$ corresponds to the
finest partition for the given data, and the analogue $b$ could be
formidably large and is not reached in our study as $b$ goes up to
64 (the same stands for $b$ up to 256). Thus direct comparison to
binning estimators when $k$ is very small cannot be drawn. For
noise-free chaotic time series, very fine partitions are sought
and this agrees with the suggestion in \citet{Kraskov04} to use
small $k$ at the order of $3$, which was also used in other
simulation studies \citep{Kreuz07,Khan07}. In
Fig.~\ref{fig:Henonneighb}a, we show for the noise-free Henon map
that ${\it I}(\tau)$ increases with decreasing $k$. \\
=============================================== \\
  Figure 9       ***  To be placed here   ***   \\
=============================================== \\
For small $n$, a large value of $k$ gives a poor estimation of the
densities and consequently of ${\it I}(\tau)$. For a fixed $k$, MI
increases with $n$ (see Fig.~\ref{fig:Henonneighb}b). Assuming a
fixed $k$ the effect of $n$ on the KNN estimator is large
similarly to the effect of $n$ on the AD estimator as there in no
convergence of MI with $n$, contrary to the fixed-bin estimators.
In agreement to the binning estimators, the MI from the KNN
estimator decreases with the noise level. Therefore the dependence
of KNN estimator on $n$ is smaller and $I(\tau)$ converges faster
to $I_{\infty}$ with $n$ (Fig.~\ref{fig:Henonneighb}c). Further,
for larger $n$ the estimation is the same regardless of the value of $k$.

The dependence of the KE estimator on the bandwidth $h_1$ is
similar to the dependence of the KNN estimator on $k$. As shown in
Fig.~\ref{fig:henonKE}a and b for the noise-free Henon map, ${\it
I}(\tau)$ increases with the decrease of $h_1$ ranging from 0.01
to 2.                                           \\
=============================================== \\
  Figure 10       ***  To be placed here   ***  \\
=============================================== \\
We note that such extremely large values of ${\it I}(\tau)$ for
very small bandwidth $h_1$ do not occur by any other estimator.
Given that the KNN estimator for $k=1$ sets an upper limit for the
estimated ${\it I}(\tau)$ on the given time series, larger ${\it
I}(\tau)$ obtained by the KE estimator are superficially overflown
estimates due to the use of an unsuitably small $h_1$ for the
given time series. This systematic bias for very small $h_1$ is
more pronounced with the addition of noise as it persists at the
same level for larger $\tau$ (see Fig.~\ref{fig:henonKE}c). For
the noisy data, ${\it I}(\tau)$ decreases and differences with
respect to the partitioning parameters are smaller, a feature we
observed also with the other estimators (see
Fig.~\ref{fig:henonKE}c and d). Also, the estimated $I(\tau)$ is
rather stable to the change of $n$.

Regarding the $9$ criteria for selecting $h_1$ (and at cases
$h_2$, see Table \ref{tab:bandwidth}), the estimated bandwidths
vary with the criterion but within a small range, e.g. for $n=256$
they are bounded in $[0.13,0.35]$ except C3 that always gives
larger bandwidths and in this case $h_1\simeq0.7$). Deviations of
the estimated bandwidths hold for larger $n$ but at smaller
magnitudes, e.g. for $n=8192$, they are bounded in $[0.03,0.18]$
and for C3 $h_1\simeq0.37$). All criteria depend on $n$ in a
similar way and estimate smaller bandwidths as $n$ increases
giving larger ${\it I}(\tau)$ (see Fig.~\ref{fig:HenonKEcrit}a for
$n=8192$). \\
=============================================== \\
  Figure 11       ***  To be placed here   ***  \\
=============================================== \\
When noise is added to the time series, the estimated ${\it
I}(\tau)$ using different bandwidth selection criteria converge
and are rather stable to the change of $n$ (see
Fig.~\ref{fig:HenonKEcrit}b).

Contrary to linear systems, for noise-free nonlinear maps, the
estimated MI does not converge to an asymptotic $I_{\infty}$ and
even for very large time series the MI values computed by
different estimators vary, as we tested for $n=10^7$. For
increasing $n$, a finer partition gives larger MI regardless of
the selected estimator. The closest approximation to the finest
partition for a large $n$ is succeeded by the KNN estimator using
a very small $k$, say $k=2$ for $n=10^7$. This turned out to be
indeed an upper bound of the estimated MI for large $n$. For the
other estimators, restrictions to the partition resolution, i.e.
smallest $h_1$ for KE and largest $b$ for the binning estimators,
bound the estimated MI to smaller values. For example, bins up to
$b=256$ for ED and EP estimators underestimate MI for $n=10^7$,
meaning that $b$ has to increase towards computationally
prohibitive large magnitudes to succeed an adequately fine
partition for this data size. In the same way, the bandwidth $h_1$
has to decrease accordingly with $n$ and for large $n$ the KE
estimator turns out to be computationally ineffective. The
presence of noise sets a lower limit to the partition resolution
and allows for an asymptotic MI value $I_{\infty}$ to which all
estimators converge with $n$ for suitably fine partition.

The results on the different estimators were only given for the
Henon map in order to facilitate comparisons, but qualitatively
similar results are obtained from the same simulations on the
Ikeda map.

\subsection{Results on nonlinear flows}
When using MI on nonlinear flows the interest is often in
extracting the lag $\tau_0$ of the first minimum of MI. We examine
the estimate of $\tau_0$ with the different MI estimators on the
Mackey-Glass system for delays $\Delta=17,30$ and $100$ that
regard increasing complexity of correlation dimension being
roughly 2,3 and 7, respectively \citep{Grassberger83}.

The simulations using the ED and EP estimators showed that the
same $\tau_0$ is estimated for all $b$, all $n$ and noise levels,
and for $\Delta=17$ and $\Delta=30$. For $\Delta=100$ we estimated
the lag for which MI levels off and there was some variation in
the selection of $\tau_0$ (see Fig.~\ref{fig:MGlassminlag}). \\
=============================================== \\
  Figure 12       ***  To be placed here   ***  \\
=============================================== \\
Although ${\it I}(\tau)$ increases with $b$, $\tau_0$ does not
vary with $b$. Moreover, the estimate of $\tau_0$ is stable with
$n$ and the addition of noise.

AD estimator is also not affected by $n$, when computing the lag
of the minimum MI $\tau_0$ in the Mackey-Glass system (see
Fig.~\ref{fig:ADMGLass}a).                      \\
=============================================== \\
  Figure 13       ***  To be placed here   ***  \\
=============================================== \\
Addition of noise does not affect the mean $\tau_0$, as shown in
Fig.~\ref{fig:ADMGLass}b. From simulations on the Mackey-Glass
system with $\Delta=100$ we observe that the mean estimated lag
that MI levels off, holds for increasing $n$ (see
Fig.~\ref{fig:ADMGLass}c) and addition of noise does not affect
it.

The estimation of $\tau_0$ using the KNN estimator on the Mackey
Glass systems varies more with $n$ and $k$ than for the binning
estimators, as shown in Fig.~\ref{fig:MGlassKNN}a and b.  \\
=============================================== \\
  Figure 14       ***  To be placed here   ***  \\
=============================================== \\
With the addition of noise, the variance of the estimated $\tau_0$
decreases with respect to $k$, and the mean is rounded to the same
integer for all $k$. For the Mackey Glass system with $\Delta=100$
we observed that there is consistency with $k$ and $n$, with MI
for all $k$ having the same shape and therefore giving the same
lag for levelling off (see Fig.~\ref{fig:MGlassKNN}c).

The mean estimated $\tau_0$ using the KE estimator on realizations
of each of the three Mackey-Glass systems is stable against
changes in the time series length and bandwidth as for the binning
estimators.

Our simulations showed that all estimators identify sufficiently
$\tau_0$ as the shape of the MI function is not affected
significantly by $n$ or by the addition of noise. ED and KE
estimators are the estimators of choice for this task, as they
give smaller variation in the estimation of $\tau_0$ compared to
the other estimators.

\section{Discussion}
\label{sec:Discussion}
MI estimators are sensitive to their free parameter, with binning
estimators (ED and EP) being the most affected. There is a loose
correspondence among the different free parameters $b$, $k$ and
$h_1$ depending also on the time series length. Thus the
differences in the performance of the estimators can be explained
to some degree by the coarseness of the partition as determined by
the free parameter. The choice of $b$ for the binning estimators
determines the bin size of the partition. The analogue of the bin
size for the KNN estimator is the size of neighborhoods given by
the number of neighbors $k$ and for the KE estimator is the size
of the efficient support of the kernel approximation given by the
bandwidth $h_1$ (and $h_2$). The simulation results have
quantified the correspondence of the different free parameters and
showed that for large time series, a suitable refined partition
can be easily accommodated by a very small $k$ or $h_1$, whereas
for the binning estimators the requirement for a very large $b$
renders the binning estimator computationally ineffective. To this
respect, the KNN estimator adapts easily to a refined partition by
setting, say, $k=2$, as does the adaptive binning estimator (AD)
that has no free parameter, whereas $h_1$ has to be further
investigated at ranges of small values.

The optimization of the parameters of the estimators is very
crucial, even more than the choice of the estimator. Therefore, we
focused on estimating the optimal free parameter of each estimator
in order to fairly evaluate the estimators. For linear systems, we
evaluated also different selection criteria for the optimal free
parameter and based on the simulation study we proposed for the
fixed-binning estimators the optimal $b$ as a function of the
autocorrelation and the time series length $n$. The parameter-free
AD estimator tends to overestimate MI compared to the other
estimators, indicating that the in-build partition algorithm of AD
terminates at a very fine partition. The KNN estimator turns out
to be the least sensitive to its free parameter. For example,
$k=2$ that gives a very fine partition does not deviate much for
smaller time series where larger $k$ are more appropriate. Our
simulation results on the linear systems have shown that the KE
estimator depends less than the binning estimators on the free
parameter for the selected ranges of $h_1$ and $b$, respectively.

For noise-free nonlinear systems, all estimators lack consistency,
i.e. the estimated MI does not converge with $n$ to an asymptotic
value. Therefore, optimal parameter cannot be derived for these
systems. The optimal parameter values found for the linear systems
tend to give conservative estimates of MI for the nonlinear
systems, for which a finer partition is required. This is
accommodated by a small $k$ in the KNN estimator. Indeed the
simulation study on the different chaotic systems has shown that
the KNN estimator has the least variance with the free parameter
$k$ than all other estimators. For noisy nonlinear systems, the MI
from all estimators converge with $n$ to an upper limit set by
noise and KNN estimator for $k=2$ turned out to reach this limit
faster.

For the computation of the lag $\tau_0$ of the first minimum of
MI, the binning estimators ED and EP as well as the KE estimator
seem to perform best. For the Mackey-Glass system, we observed
that although $I(\tau)$ may vary with the free parameter of the
estimator and $n$, $\tau_0$ is rather stable. The addition of
noise does not seem to effect the estimation of $\tau_0$.

The KE estimator has the highest computational cost and the
fixed-binning estimators become computationally intractable when
$b$ has to be very large, as for long chaotic time series. On the
other hand, the KNN estimator is rather fast for long time series
that require small $k$ for which neighbor search is faster. The
computation efficiency of the AD estimator is comparable to that
of KNN and these two estimators seem to be the most appropriate
for all practical purposes in terms of computational efficiency,
parameter selection (small $k$ for KNN and no free parameter for
AD) and accuracy of estimation (with KNN scoring better than AD).

We note that the consistency of estimators of MI on linear systems is
not indicative of the behavior of the estimators on nonlinear systems.
Although consistency of estimators is claimed in some recent works,
this might be due to the use of only linear systems or noisy real data,
such as EEG.

\section{Acknowledgments}
This research project is implemented within the framework of the
"Reinforcement Programme of Human Research Manpower" (PENED) and
is co-financed at $90\%$ jointly by E.U.-European Social Fund
($75\%$) and the Greek Ministry of Development-GSRT ($25\%$) and
at $10\%$ by Rikshospitalet, Norway.

\clearpage

\begin{table} [h!]
\caption{The simulation systems and their parameters. The input
white noise for the linear systems (rows 4 to 7) and the
observational white noise for the nonlinear systems (rows 8 to 10)
have zero mean and standard deviation one. Gamma noise is skewed
with $\gamma=0.5$.The parameter notations are $\mu$ for the mean,
$\sigma$ for the standard deviation and $\gamma$ for the skewness
coefficient, $\varphi$ for the coefficient of the autoregressive
part for AR(1) and ARMA(1,1) and $\vartheta$ for the coefficient
of the moving average part of ARMA(1,1), $\tau_s$ for the sampling
time and $\delta$ for the discretization time for the Mackey-Glass
system. The noise levels considered for the nonlinear systems are
$20\%$ and $40\%$.}
\begin{tabular}{|l|l|l|c|} \hline
{\em Systems} & {\em Parameters} & {\em Noise} \\ \hline
Gaussian white noise  & $\mu = 0, \sigma=1$ & \\
Gamma white noise     & $\mu = 0, \sigma=1,\gamma=0.5$ & \\\hline
AR(1) & $\varphi=0.5,0.9,-0.5,-0.9$ & Gaussian \\
AR(1) & $\varphi=0.5,0.9,-0.5,-0.9$ & Gamma \\\hline
ARMA(1,1) & $\varphi=0.9,\vartheta=0.6$ \& $\varphi=0.7,\vartheta=0.3$ & Gaussian \\
ARMA(1,1)& $\varphi=0.7,\vartheta=0.3$ \&
$\varphi=0.3,\vartheta=0.1$ & Gamma\\\hline
Henon &  $a=1.4,b=0.3$  & Gaussian \\
Ikeda &  $a=1.0,b=0.9,\kappa=0.4,\eta=6.0$  & Gaussian \\
Mackey-Glass & $\Delta=17,30,100$, $\tau_s=17$, $\delta=0.1$ &
Gaussian \\\hline
\end{tabular}
\label{tab:systems}
\end{table}

\clearpage

\begin{table}
\caption{Criteria for the selection of the number of bins. The
parameters in the expressions are the time series length $n$, the
standard deviation $s$, the interquartile range IQR, the range of
the data $R$ and the standardized skewness $\gamma_2$ as defined
in \citep{Doane76}. The exact expressions for criteria H8 and H9
can be found in the in the corresponding references, given in the
third column.}.
\begin{tabular}{| c | l | c |} \hline
Criteria    &    Number of bins       &   Reference       \\\hline
H1 & $1+\log_2{n}$                    & \citep{Sturge26}  \\\hline
H2 & $1.87(n-1)^{0.4}$                & \citep{Bendat66}  \\\hline
H3 & $1+\log_2{n}+\log_2\gamma_2$     & \citep{Doane76}   \\\hline
H4 & $\sqrt{n}$                       & \citep{Tukey77}   \\\hline
H5 & $\frac{Rn^{1/3}}{3.49s}$         & \citep{Scott79}   \\\hline
H6 & $\frac{Rn^{1/3}}{{2(IQR)}}$      & \citep{Freedman81}\\\hline
H7 & $\sqrt[3]{2n}$                   & \citep{Terrell85} \\\hline
H8 & min. of stochastic complexity    & \citep{Rissanen92}\\\hline
H9 & mode of log of marginal posterior pdf &
\citep{Knuth06}\\\hline H10 & $\sqrt{n/5}$   & \citep{Cochran54}
\\\hline
\end{tabular}
\label{tab:binnumber}
\end{table}

\clearpage

\begin{table}
\caption{Criteria for the selection of bandwidths for one ($h_1$)
and two ($h_2$) dimensions. The parameters in the expressions are
$a=1.8-r(1)$ if $n<200$ and $a=1.5$ if $n \geq 200$, where $r(1)$
is the autocorrelation at lag $1$, $R=1/2{\sqrt{\pi}}$, $s$ is the
standard deviation and $\mbox{IQR}$ is the interquartile range of
the data. The exact expressions for the last four criteria can be
found in the corresponding references, given in the third column.}
\begin{tabular}{|c|l|l|l|} \hline
Criteria   &   $h_1$    &    $h_2$   &       Reference    \\\hline
C1 & $(4/3n)^{1/5}$  & $(1/n)^{1/6}$ & \citep{Silverman86}\\\hline
C2 & $(4/3n)^{1/5}$  & $(4/5n)^{1/6}$& \citep{Silverman86}\\\hline
C3 & $1.06an^{-1/5}$ & $an^{-1/6}$   & \citep{Harrold01}  \\\hline
C4 &     & $h_1$ & \citep{Silverman86} \\ \cline{1-1}  \cline{3-3}
C5&\raisebox{2.5mm}[0pt]{$(\frac{8{\sqrt{\pi}}R}{3n})^{1/5}\min(s,\frac{\mbox{IQR}}{1.349})$}
& $\sqrt{2}h_1$ & \citep{Wand95} \\ \hline C6 &   &  $h_1$   &
\\  \cline{1-1}   \cline{3-3} C7 & \raisebox{2.5mm}[0pt]{L-stage
direct plug-in} & $\sqrt{2}h_1$ &
\raisebox{2mm}[0pt]{\citep{Wand95}} \\ \hline C8 &   & $h_1$  & \\
\cline{1-1}   \cline{3-3} C9
&\raisebox{2.5mm}[0pt]{Solve-the-equation plug-in} & $\sqrt{2}h_1$
& \raisebox{2mm}[0pt]{\citep{Sheather91}} \\\hline
\end{tabular}
\label{tab:bandwidth}
\end{table}

\clearpage

\begin{table}
\caption{Ranking and score $S$ of the $5$ criteria for $b$ scoring
lowest for the \emph{ED} and \emph{EP} estimators for varying
lengths of time series from white noise and linear systems.}
\begin{tabular}{|c|c|c|c|} \hline
Criteria & $S$ for \emph{ED} & Criteria & $S$ for \emph{EP}
\\ \hline
H11    & 0.25     &   H9      &   0.61                  \\ \hline
H7     & 0.94     &   H7      &   0.62                  \\ \hline
H8     & 1.16     &   H8      &   0.65                  \\ \hline
H5     & 1.20     &   H10     &   0.67                  \\ \hline
H9     & 1.41     &   H11     &   0.72                  \\ \hline
\end{tabular}
 \label{tab:optCriteria}
\end{table}

\clearpage

\section*{Figures captions}
Figure 1: Mean $I(1)$ as a function of $b$ from 1000 realizations
of AR(1) with coefficient $\varphi=r(1)=0.5$ and additive Gaussian
noise for (a) the ED and (b) the EP estimator and for data sizes
as given in the legend.

\medskip

Figure 2: (a) Optimal number of bins for different $n$ for AR(1)
systems with lag one autocorrelation $r(1)$ as in the legend. (b)
Graph of the optimal $b$ for a range of $\log_2(n)$ and $r(1)$.
The results in both panels regard the ED estimator.

\medskip

Figure 3: Mean estimated MI with AD estimator as a function of $n$
from 1000 realizations of (a) normal white noise and (b) AR(1),
with $r=0.5$ and normal input white noise.

\medskip

Figure 4: (a) Mean estimated MI with the KNN estimator as a function
of $n$ from 1000 realizations of AR(1) with $r(1)=0.5$ and normal
input white noise for $k$ as in the legend. The dotted line stands
for $I_{\infty}$. (b) Graph of the optimal $k$ for a range of
$\log_2(n)$ and $r(1)$.

\medskip

Figure 5: Mean estimated MI with the KE estimator as a function of
$h_1$ from 1000 realizations of $AR(1)$ with $r(1)=0.5$ and normal
input white noise for (a) $h_1=h_2$, and (b) $h_1=\sqrt2{h_2}$,
and $n$ as in the legend. (c) Graph of the optimal $h_1$ for a
range of $\log_2(n)$ and $r(1)$.

\medskip

Figure 6: (a) Mean estimated MI vs $n$ for the estimators given in
the legend from simulations on AR(1) with $r(1)=0.5$ and normal
input white noise. For each estimator the optimal free parameter
is considered, i.e. H11 for ED, H9 for EP, $k=2$ for KNN and C1
for KE. (b) As (a) but for $r(1)=0.9$.

\medskip

Figure 7: (a) Optimal number of bins $b$ as a function of the time
series length $n$ for the ED estimator from 1000 realizations of
the Henon map for different lags, as given in the legend. (b) Same
graph as in (a) but for the Henon map with $20\%$ additive noise.
In both plots the dotted lines give the optimal number of bins $b$
from the suggested criterion in (\ref{eq:numbins}) assuming r(1)=0
and r(1)=1, as given in the plot. (c) Mean estimated MI with ED
estimator as a function of $\tau$ from 1000 realizations of the
Henon map for $b=32$, and $n$ as in the legend. (d) As (c) but
for Henon map with $20\%$ additive noise.

\medskip

Figure 8: Mean estimated MI with AD estimator as a function of
$\tau$ from 1000 realizations of the Henon map with no noise in (a)
and with $20 \%$ noise in (b).

\medskip

Figure 9: (a) Mean estimated MI with KNN estimator as a function of
$\tau$ from $1000$ realizations of the Henon map, for $n=256$
and $k$ as in the legend. (b) As in (a) but for $k=2$ and $n$ as in the legend.
(c) As in (a) but for Henon map with $20\%$ additive noise.

\medskip

Figure 10: (a) Mean estimated MI with KE estimator as a function of
$\tau$ from $1000$ realizations of the Henon map, for $n=512$
and bandwidths as in the legend. (b) As in (a)
but for $n=4096$ and bandwidths as in the legend. (c) and (d)
are the same as (a) and (b) respectively but for the Henon map with
$20 \%$ additive noise.

\medskip

Figure 11: (a) Mean estimated MI with KE estimator as a function of
$\tau$ from $1000$ realizations of the noise-free Henon map with $n=8192$
and nine bandwidth selection criteria as given in the legend. (b)
As (a) but for $20\%$ additive noise.

\medskip

Figure 12: (a) Mean estimated $\tau_0$ and standard deviation as
error bar as a function of $b$ from all time series lengths using
the ED estimator on the Mackey-Glass system with
$\Delta=17,30,100$, as given in the legend. (b) As in (a) but for
the EP estimator.

\medskip

Figure 13: (a) Mean estimated $\tau_0$ and standard deviation as
error bar as a function of $n$ using the AD estimator on 1000
realizations of the Mackey-Glass system with $\Delta=17,30$, as
given in the legend. (b) As in (a) (without standard deviations)
for additive noise with levels $20, 40 \%$, as given in the
legend. (c) Mean estimated MI with the AD estimator as a function
of $\tau$ from 1000 realizations of the Mackey-Glass with
$\Delta=100$, for $n$ as in the legend.

\medskip

Figure 14: Mean estimated $\tau_0$ as a function of $n$ using the
KNN estimator on 1000 realizations of the Mackey-Glass system with
(a) $\Delta = 17$, and (b) $\Delta = 30$. (c) Mean estimated MI
with the KNN estimator as a function of $\tau$ from 1000
realizations of the Mackey-Glass with $\Delta=100$, for $k$ as in
the legend and $n=2048$.

\clearpage

\clearpage

\clearpage

\clearpage

\clearpage

\clearpage

\begin{figure*}
\centerline{\hbox{
\includegraphics[height=9cm,keepaspectratio]{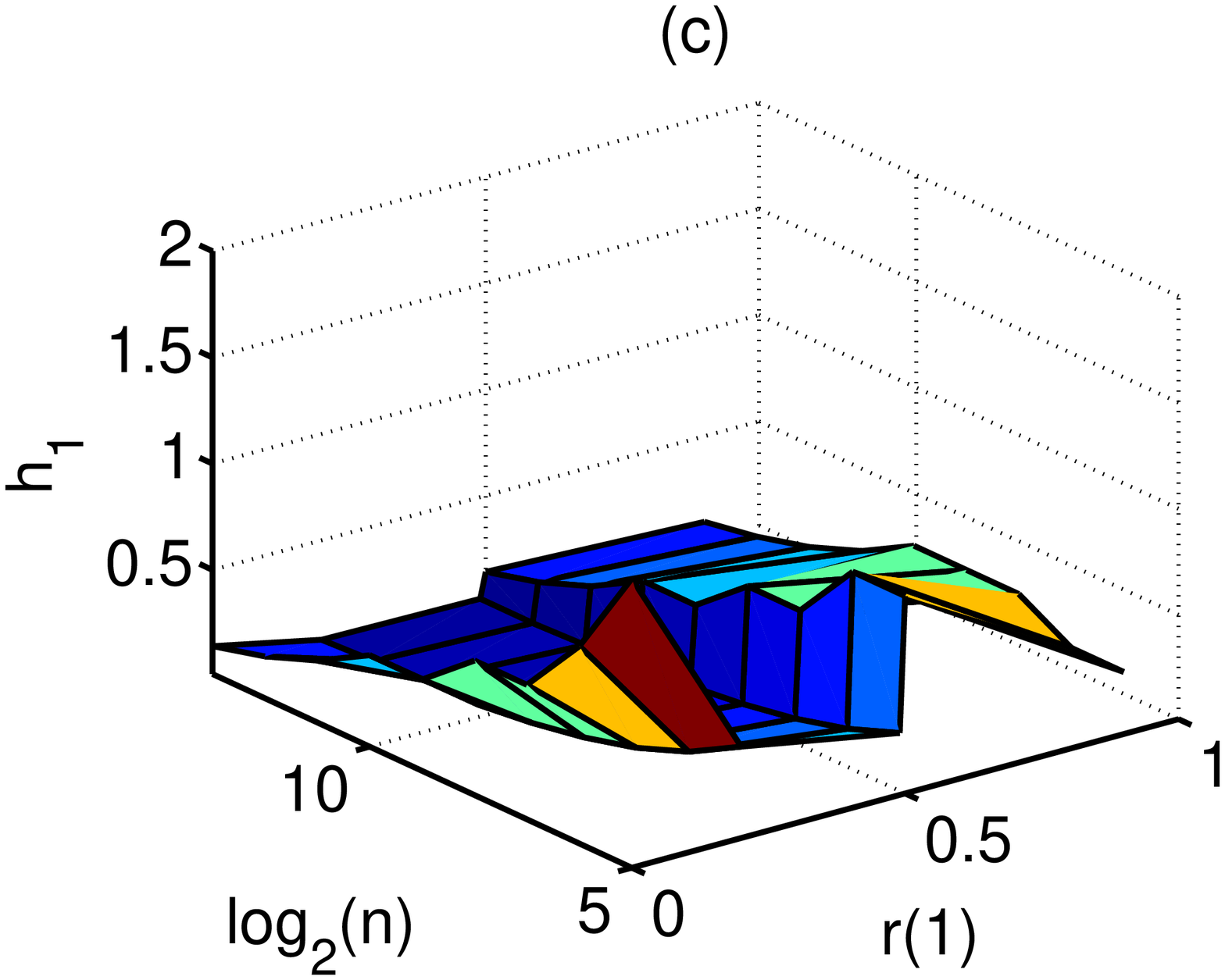}}}
\end{figure*}
\centering{Figure 5c: A. Papana}

\clearpage
\begin{figure}
\centerline{\hbox{
\includegraphics[height=9cm,keepaspectratio]{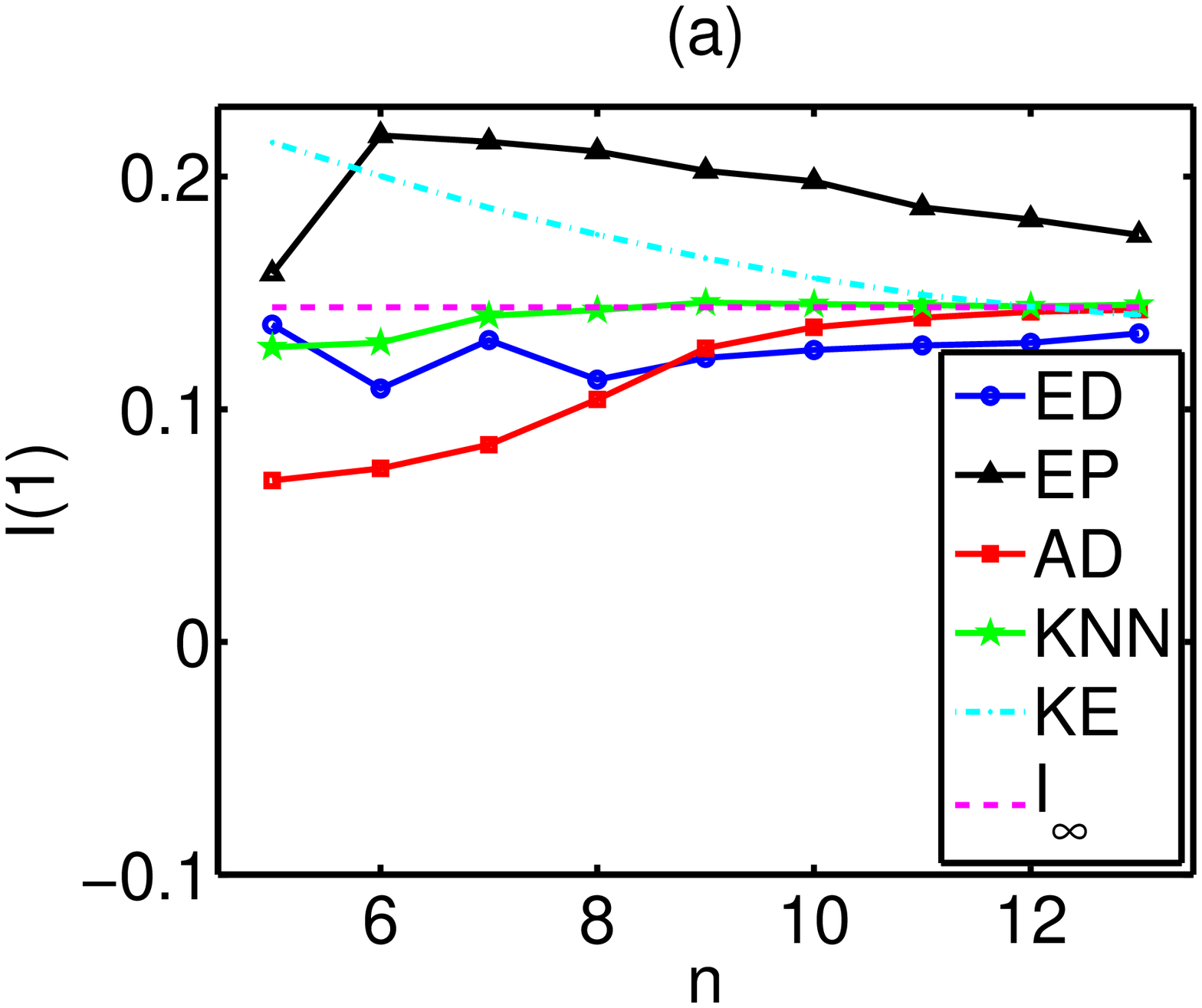}}}
\centerline{\hbox{
\includegraphics[height=9cm,keepaspectratio]{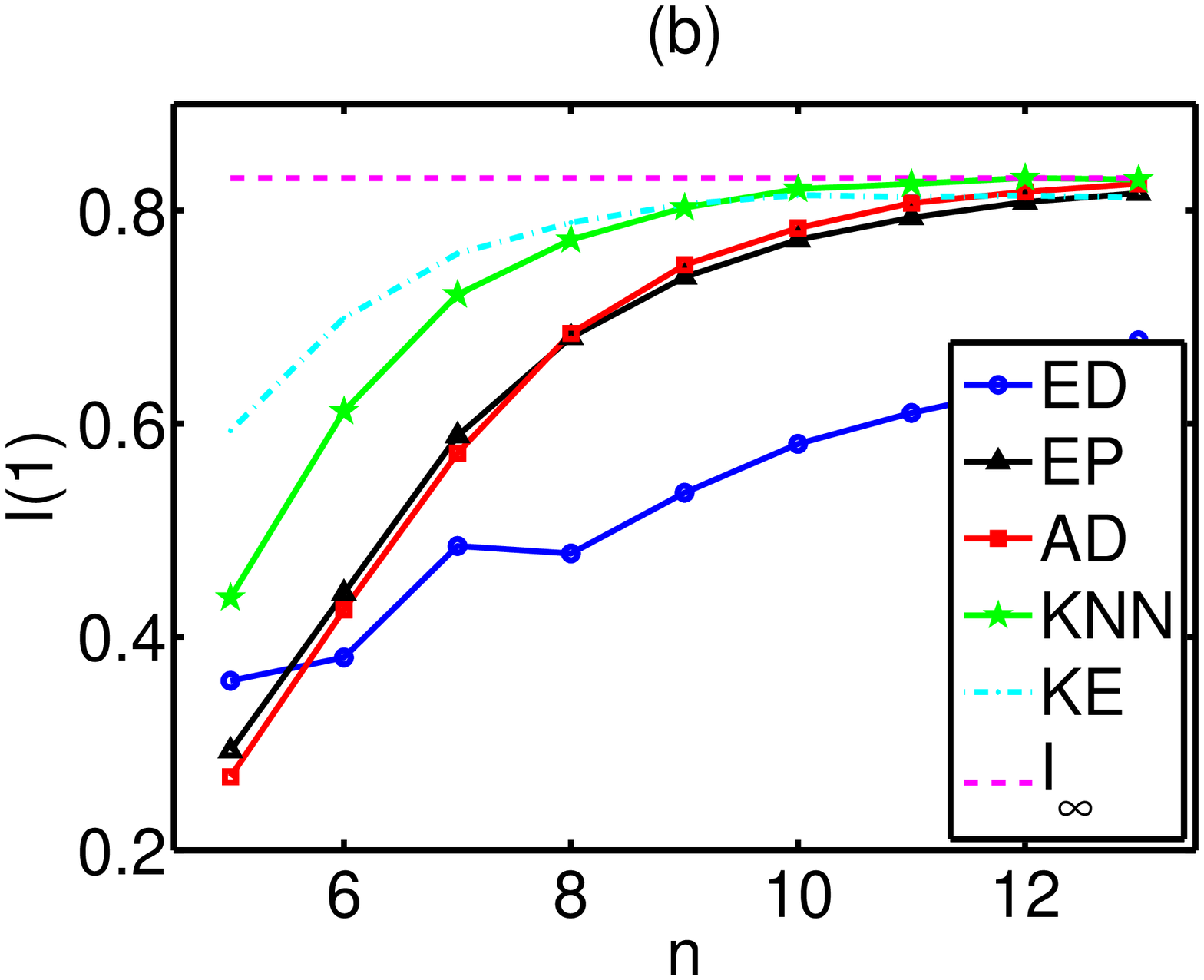}}}
\caption{A. Papana}
 \label{fig:alllinear}
\end{figure}

\clearpage
\begin{figure}
\centerline{\hbox{
\includegraphics[height=9cm,keepaspectratio]{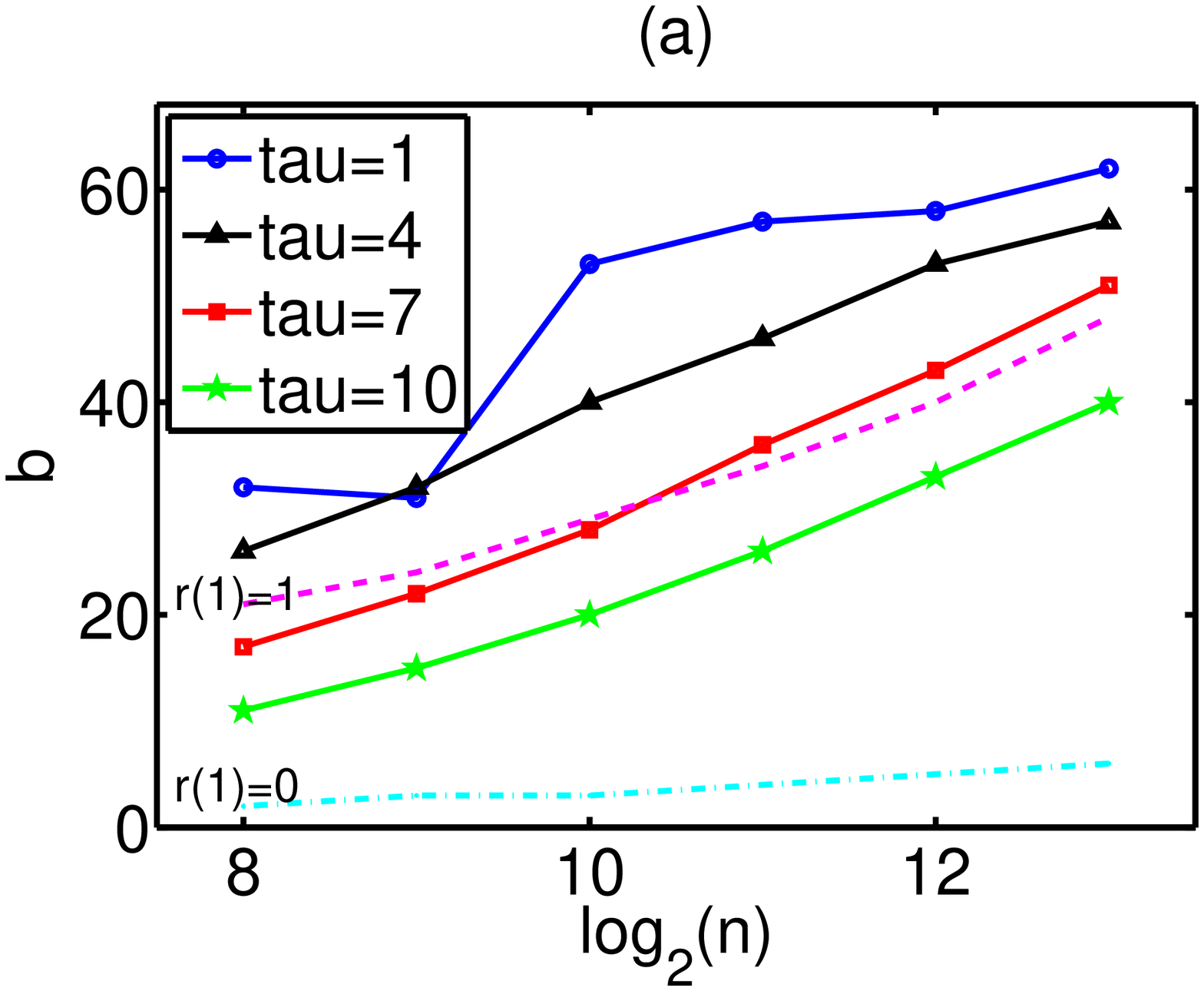}}}
\centerline{\hbox{
\includegraphics[height=9cm,keepaspectratio]{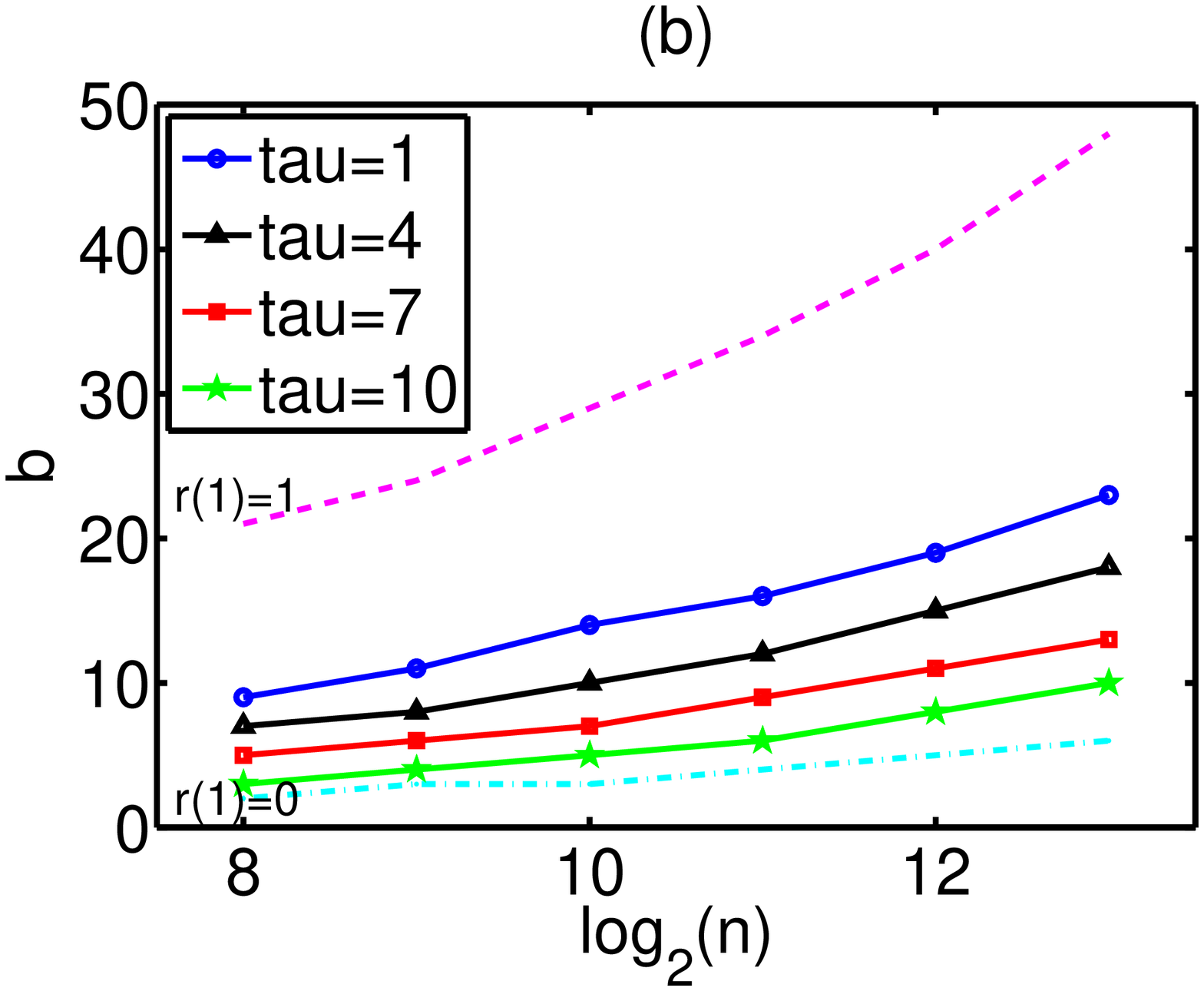}}}
\caption{A. Papana} \label{fig:henonoptbin}
\end{figure}

\clearpage
\begin{figure*}[h!]
\centerline{\hbox{
\includegraphics[height=9cm,keepaspectratio]{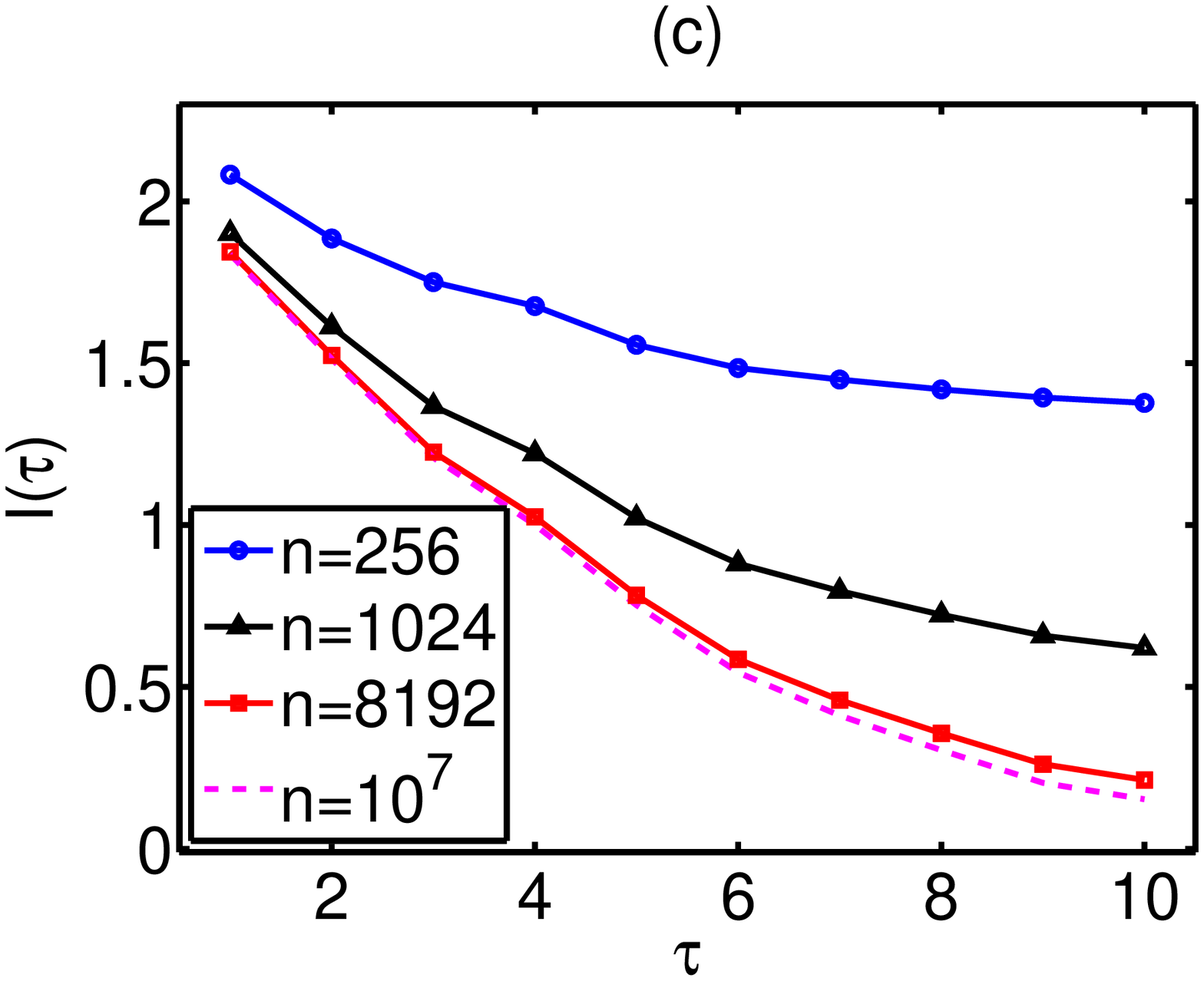}}}
\centerline{\hbox{
\includegraphics[height=9cm,keepaspectratio]{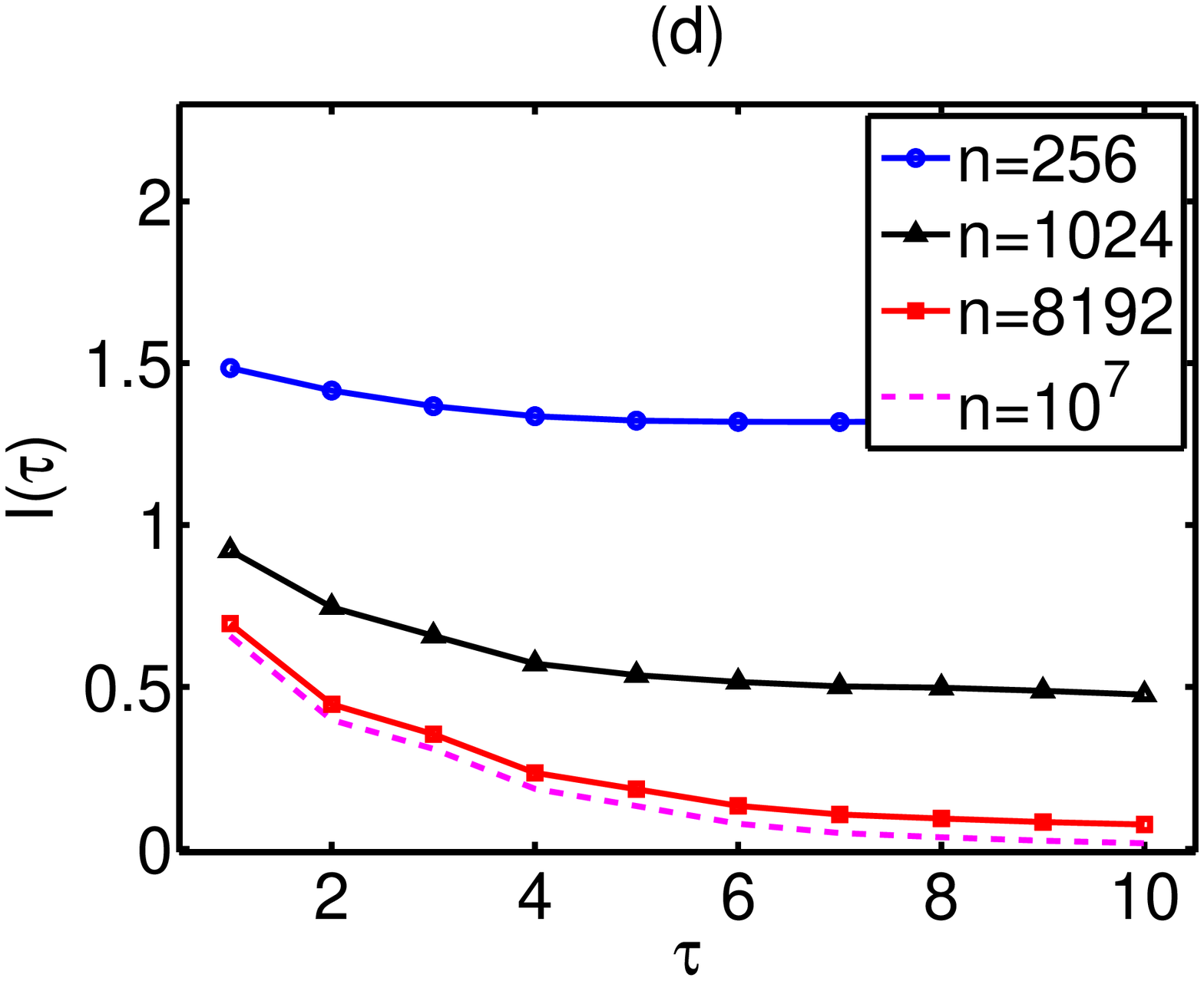}}}
\end{figure*}
\centering{Figure 7c and d: A. Papana}

\clearpage
\begin{figure}
\centerline{\hbox{
\includegraphics[height=9cm,keepaspectratio]{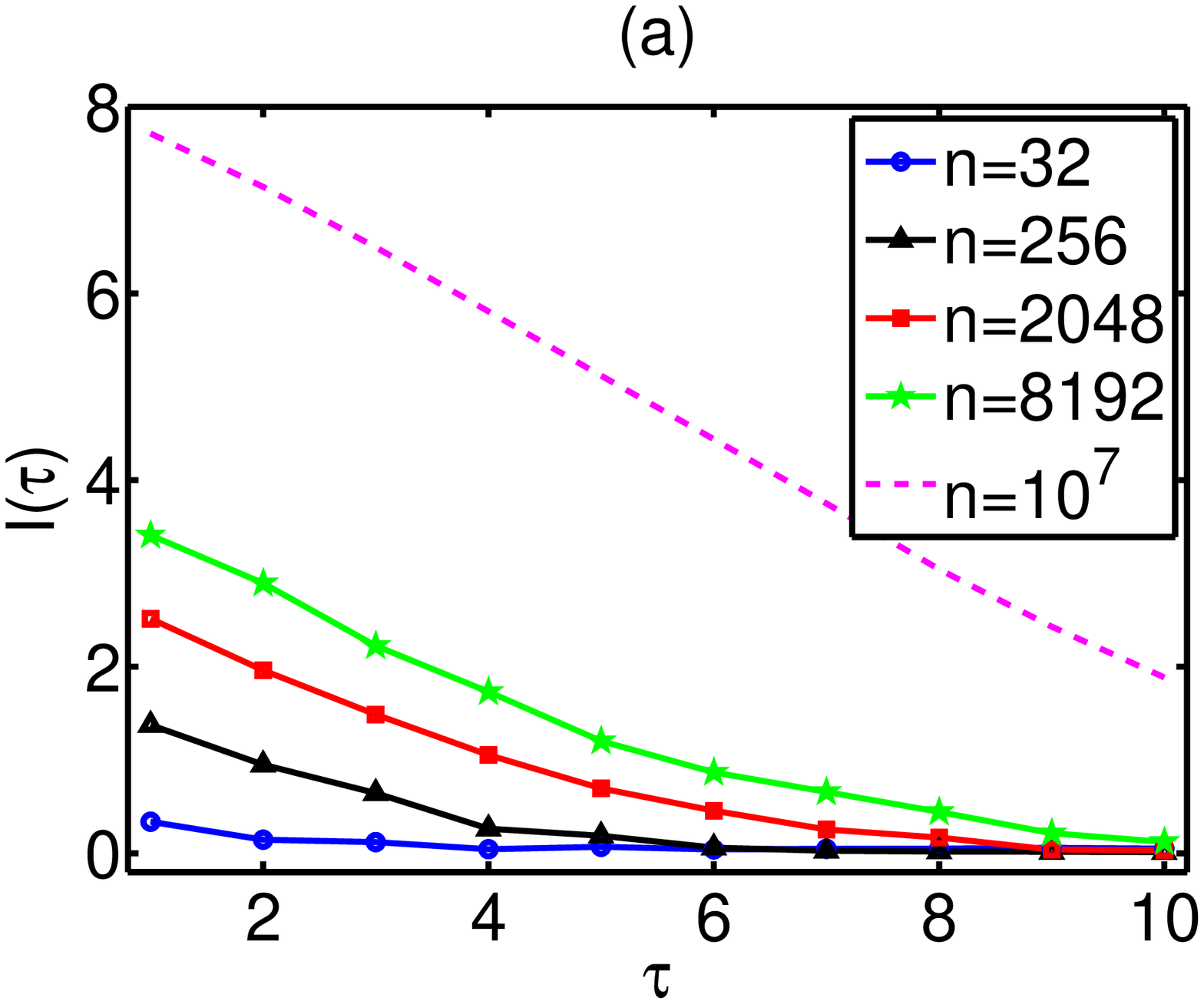}}}
\centerline{\hbox{
\includegraphics[height=9cm,keepaspectratio]{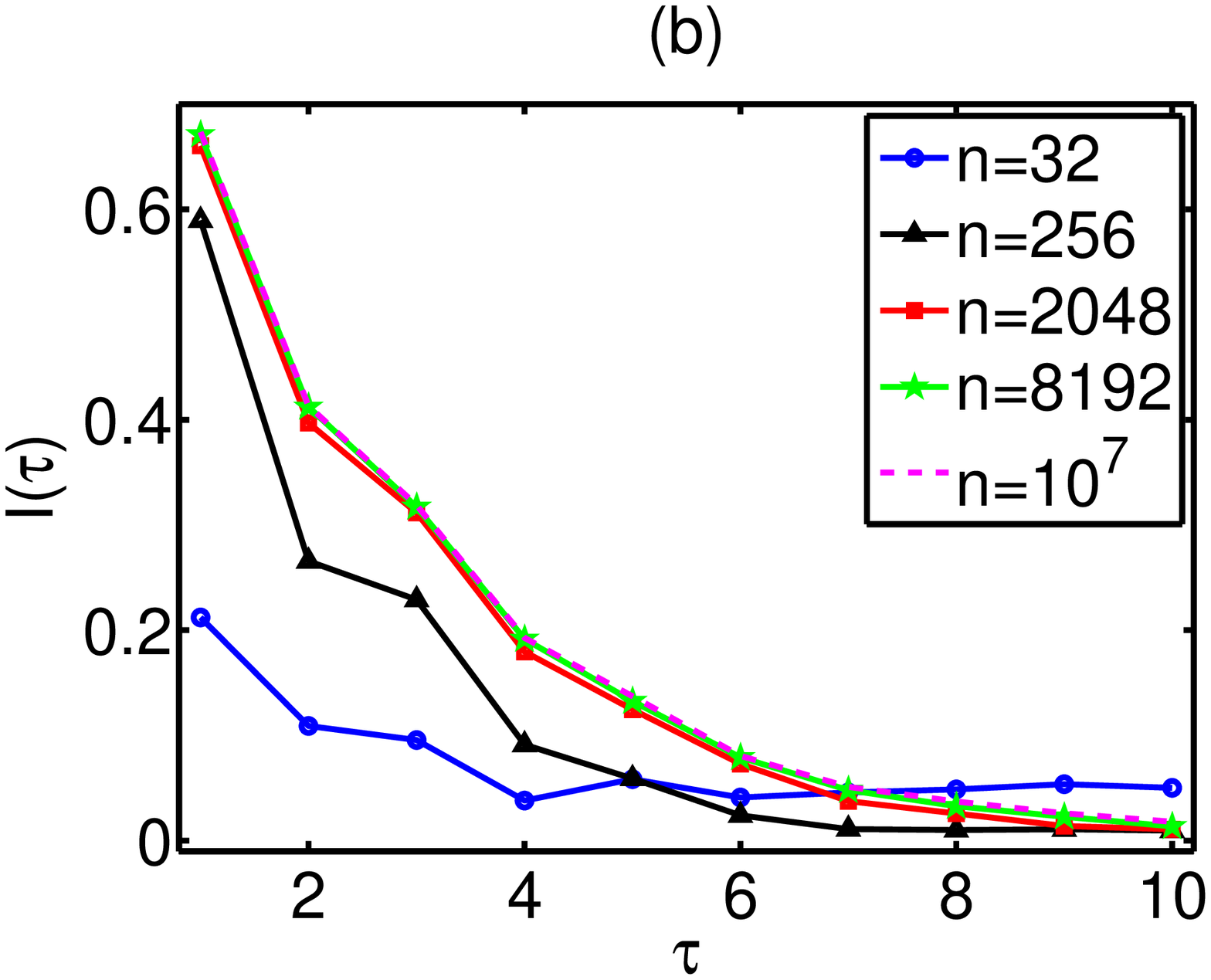}}}
\caption{A. Papana} \label{fig:MIAD}
\end{figure}

\clearpage
\medskip
\begin{figure}
\centerline{\hbox{
\includegraphics[height=9cm,keepaspectratio]{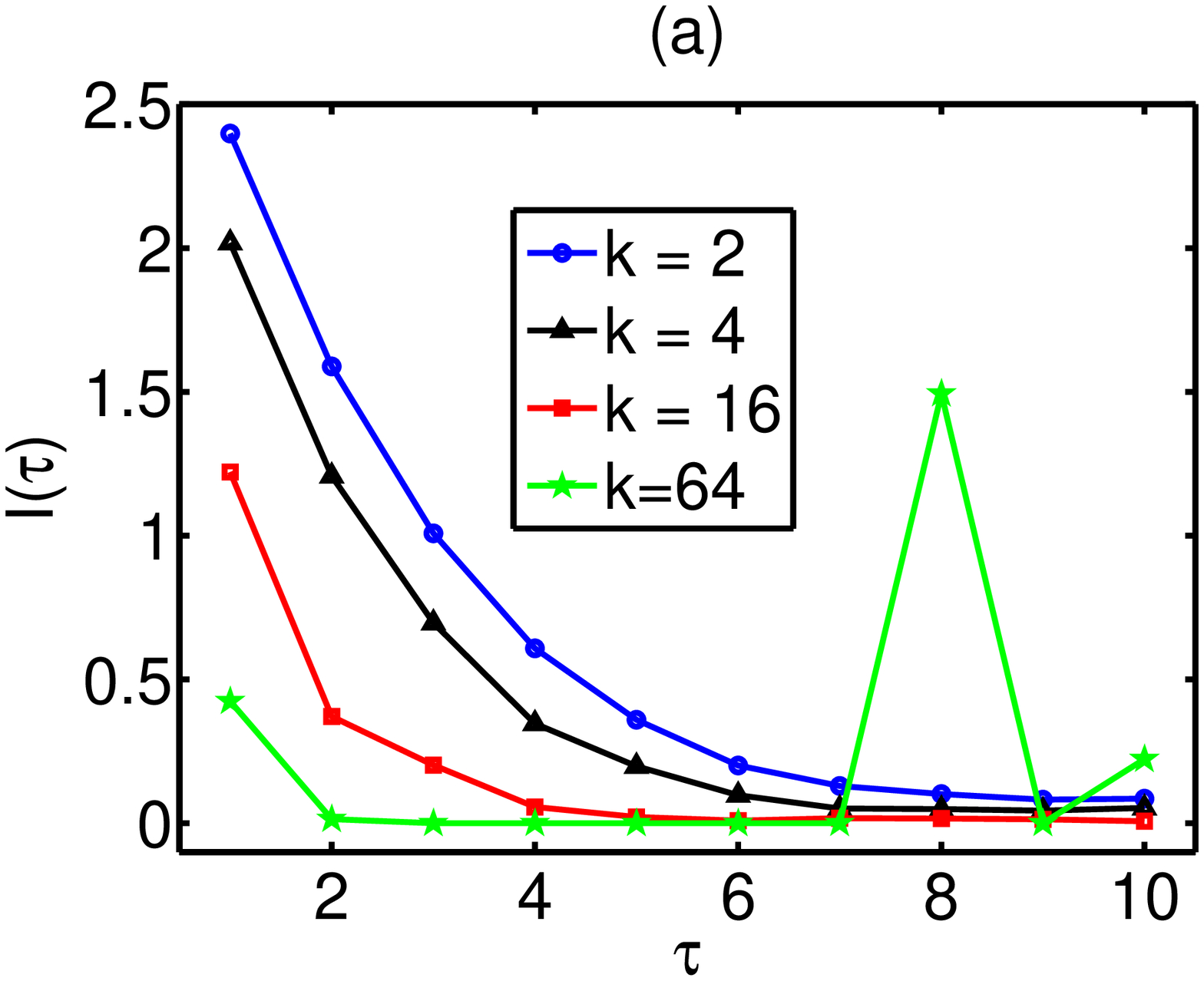}}}
\centerline{\hbox{
\includegraphics[height=9cm,keepaspectratio]{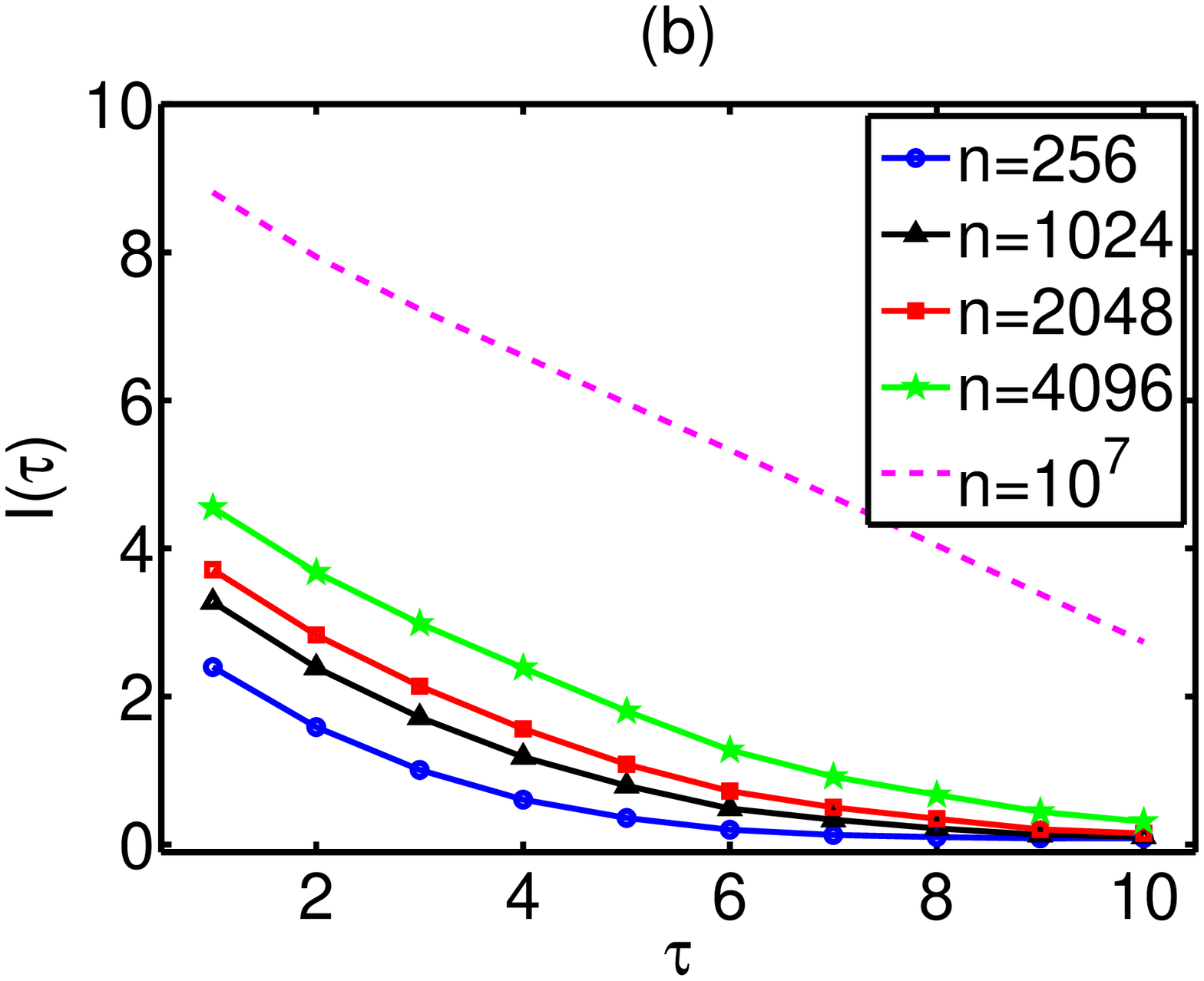}}}
\caption{A. Papana} \label{fig:Henonneighb}
\end{figure}

\clearpage

\begin{figure*}
\centerline{\hbox{
\includegraphics[height=9cm,keepaspectratio]{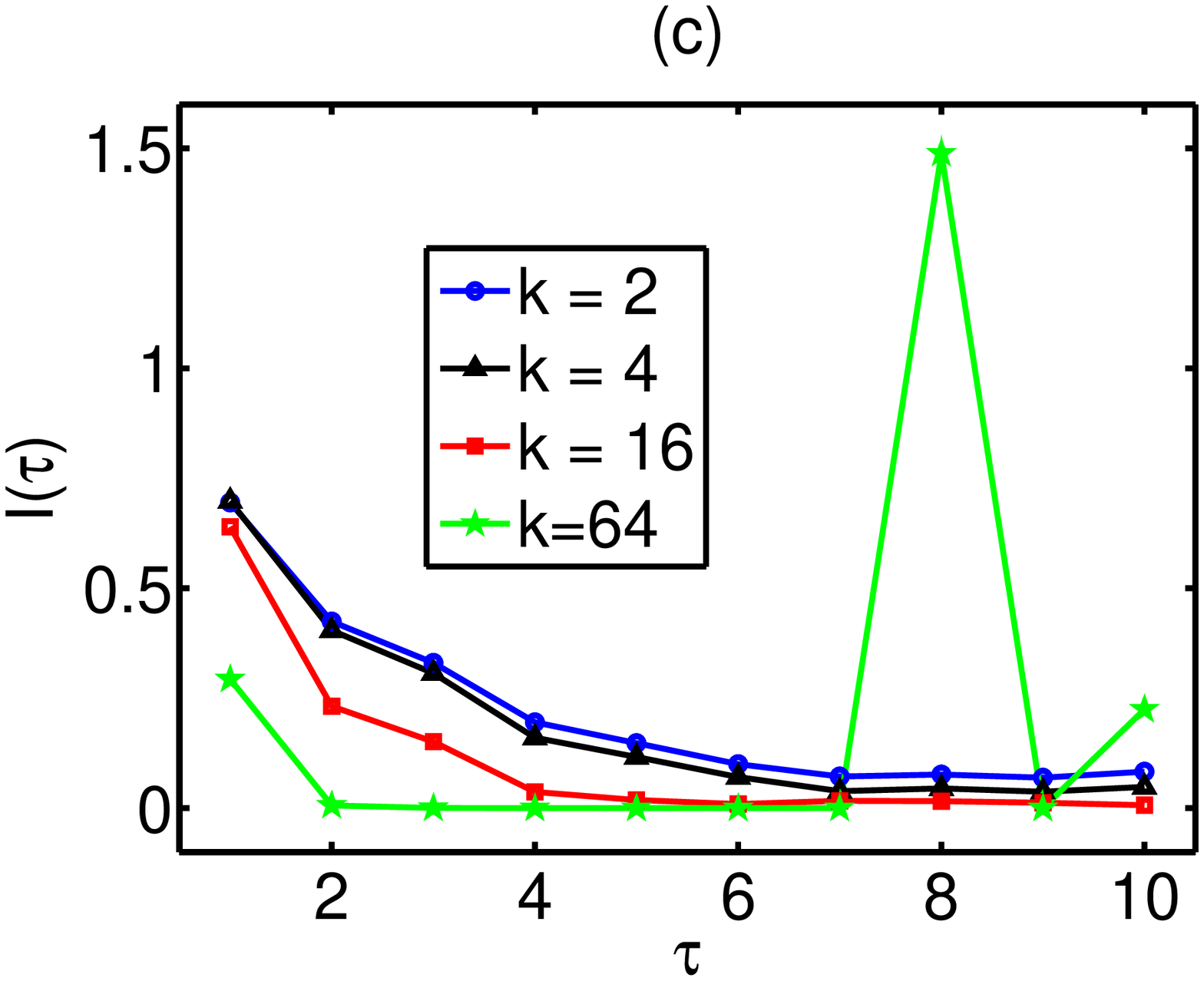}}}
\centering{Figure 9c: A. Papana}
\end{figure*}

\clearpage
\begin{figure}
\centerline{\hbox{
\includegraphics[height=9cm,keepaspectratio]{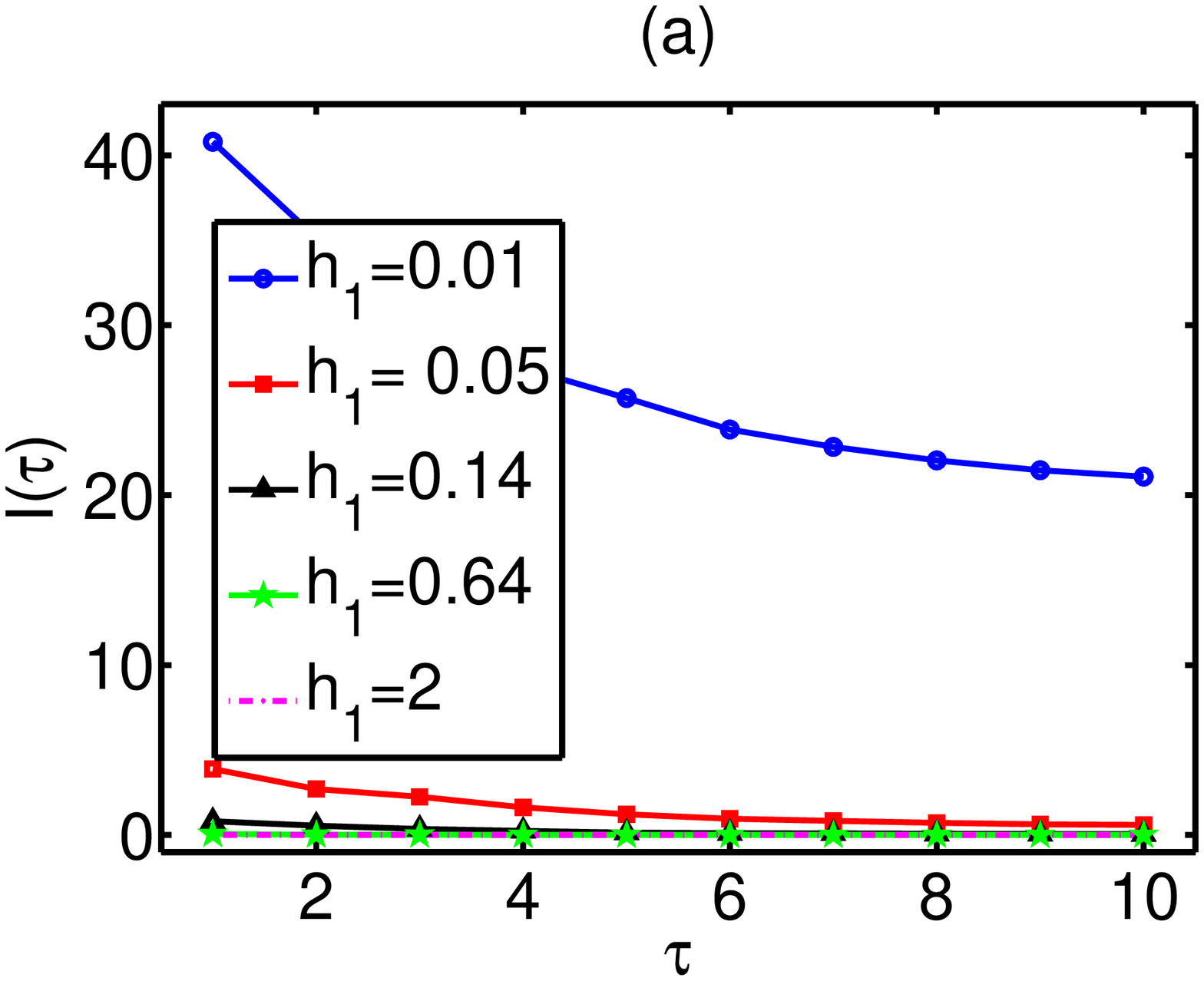}}}
\centerline{\hbox{
\includegraphics[height=9cm,keepaspectratio]{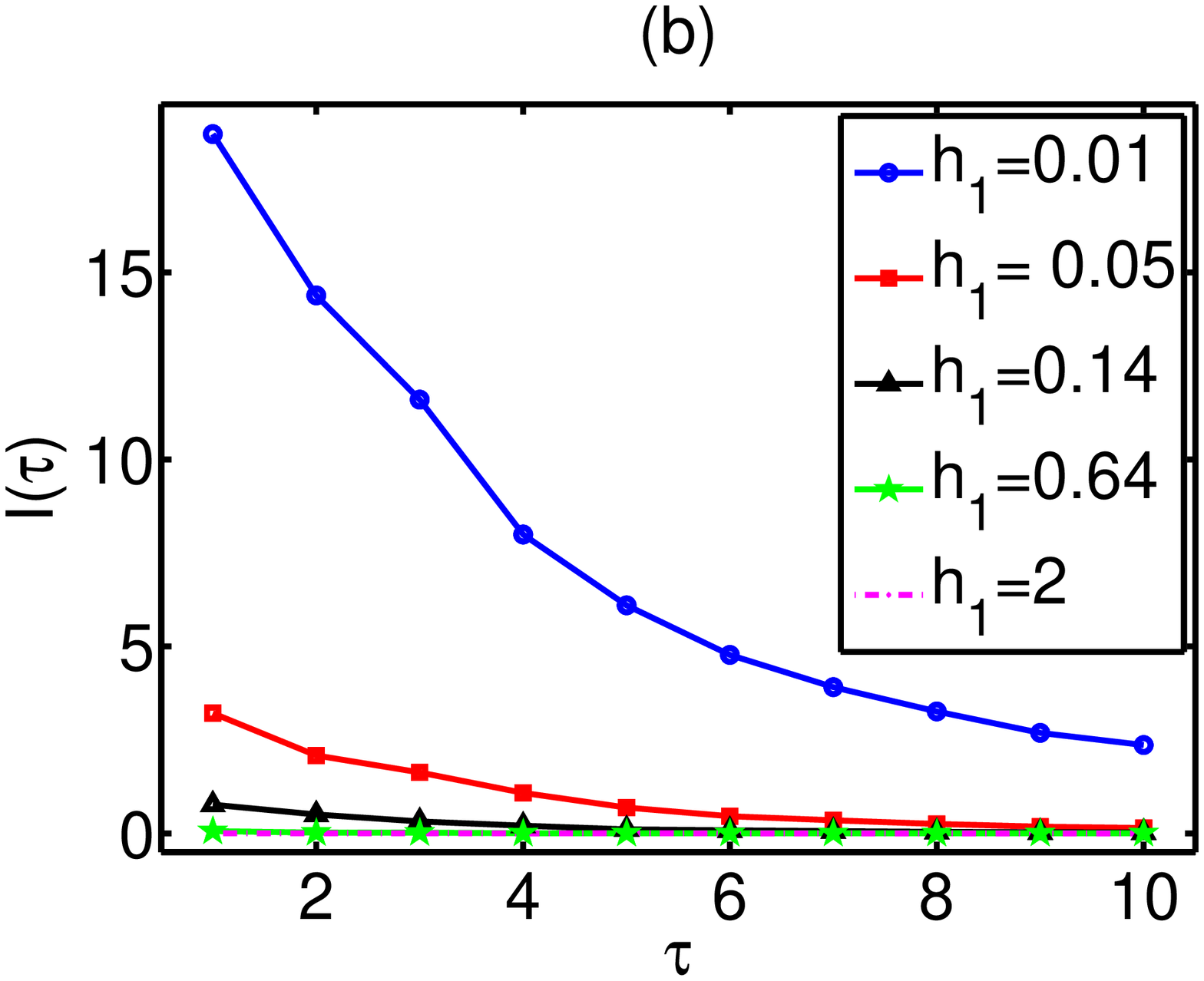}}}
 \caption{A. Papana} \label{fig:henonKE}
\end{figure}

\clearpage

\begin{figure*}
\centerline{\hbox{
\includegraphics[height=9cm,keepaspectratio]{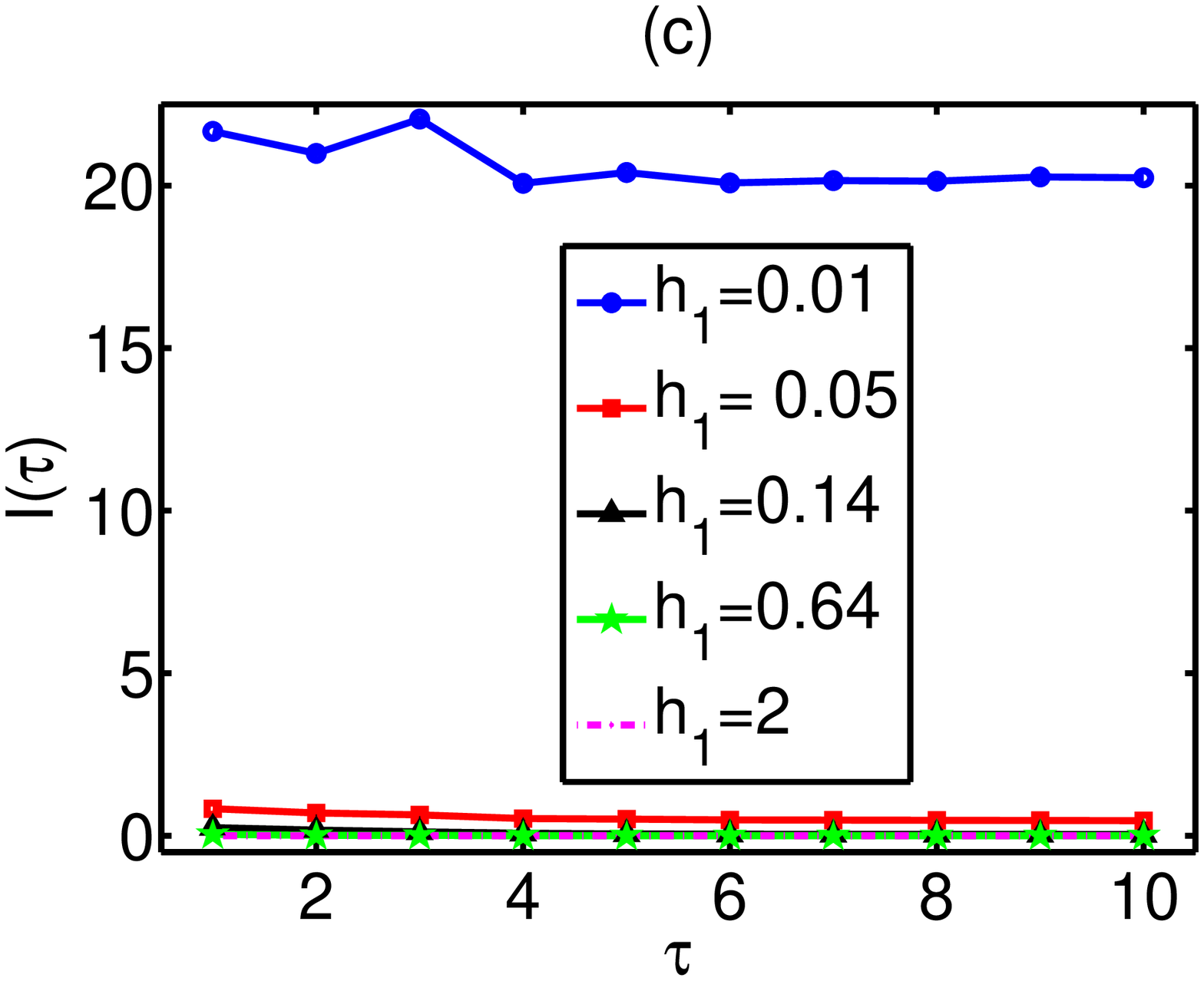}}}
 \centerline{\hbox{
\includegraphics[height=9cm,keepaspectratio]{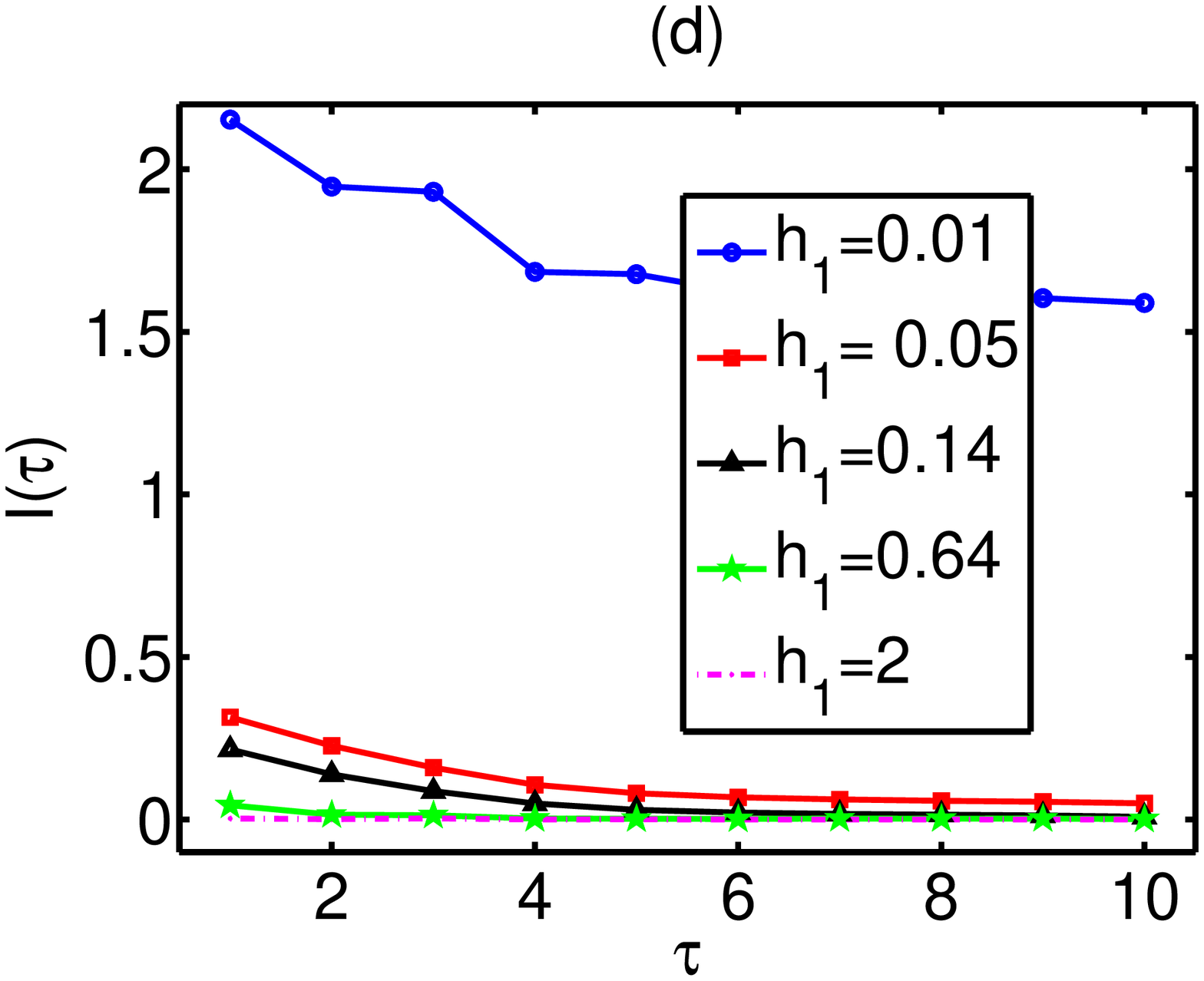}}}
\centering{Figure 10c and d: A. Papana}
\end{figure*}

\clearpage
\begin{figure}
\centerline{\hbox{
\includegraphics[height=9cm,keepaspectratio]{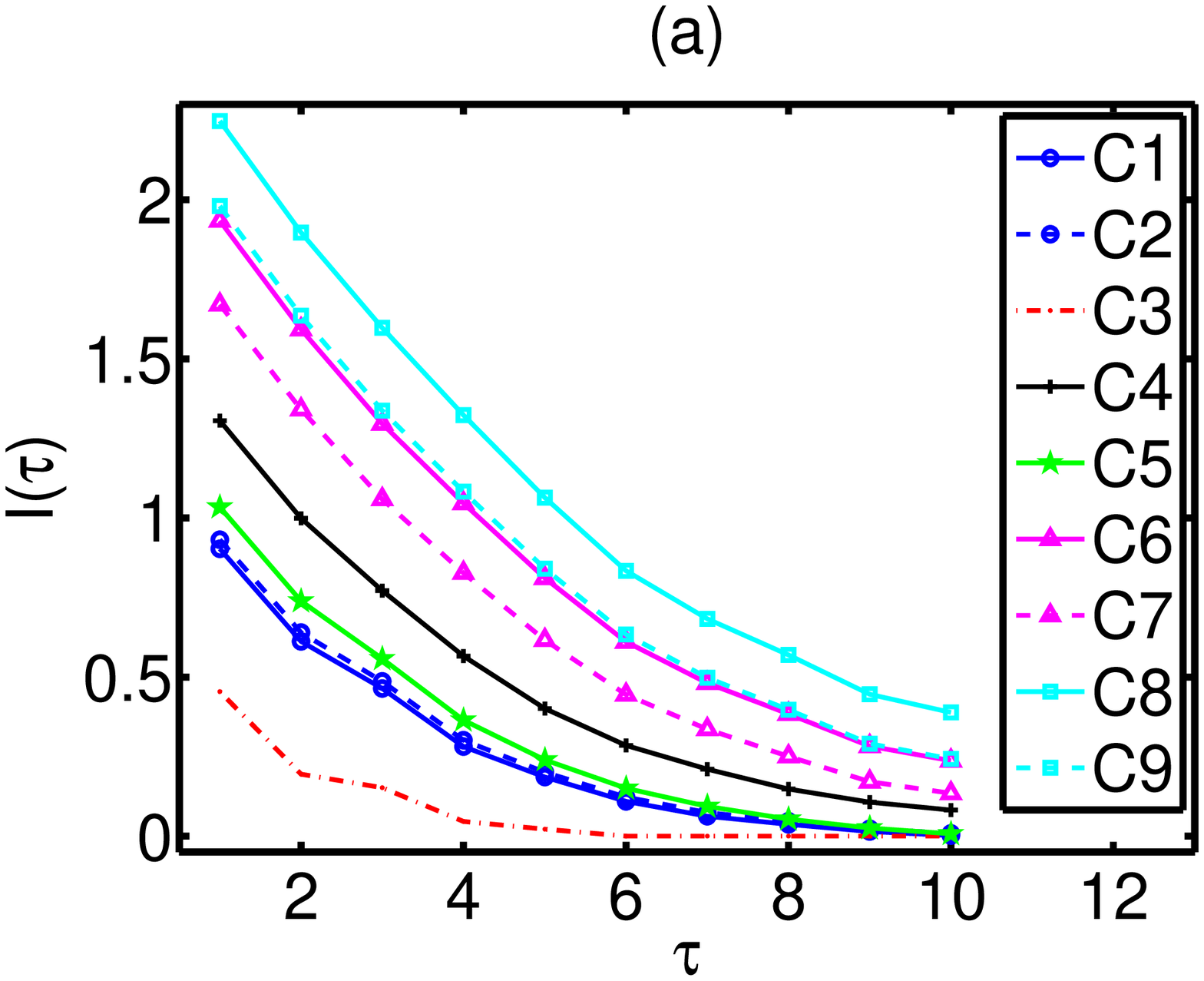}}}
\centerline{\hbox{
\includegraphics[height=9cm,keepaspectratio]{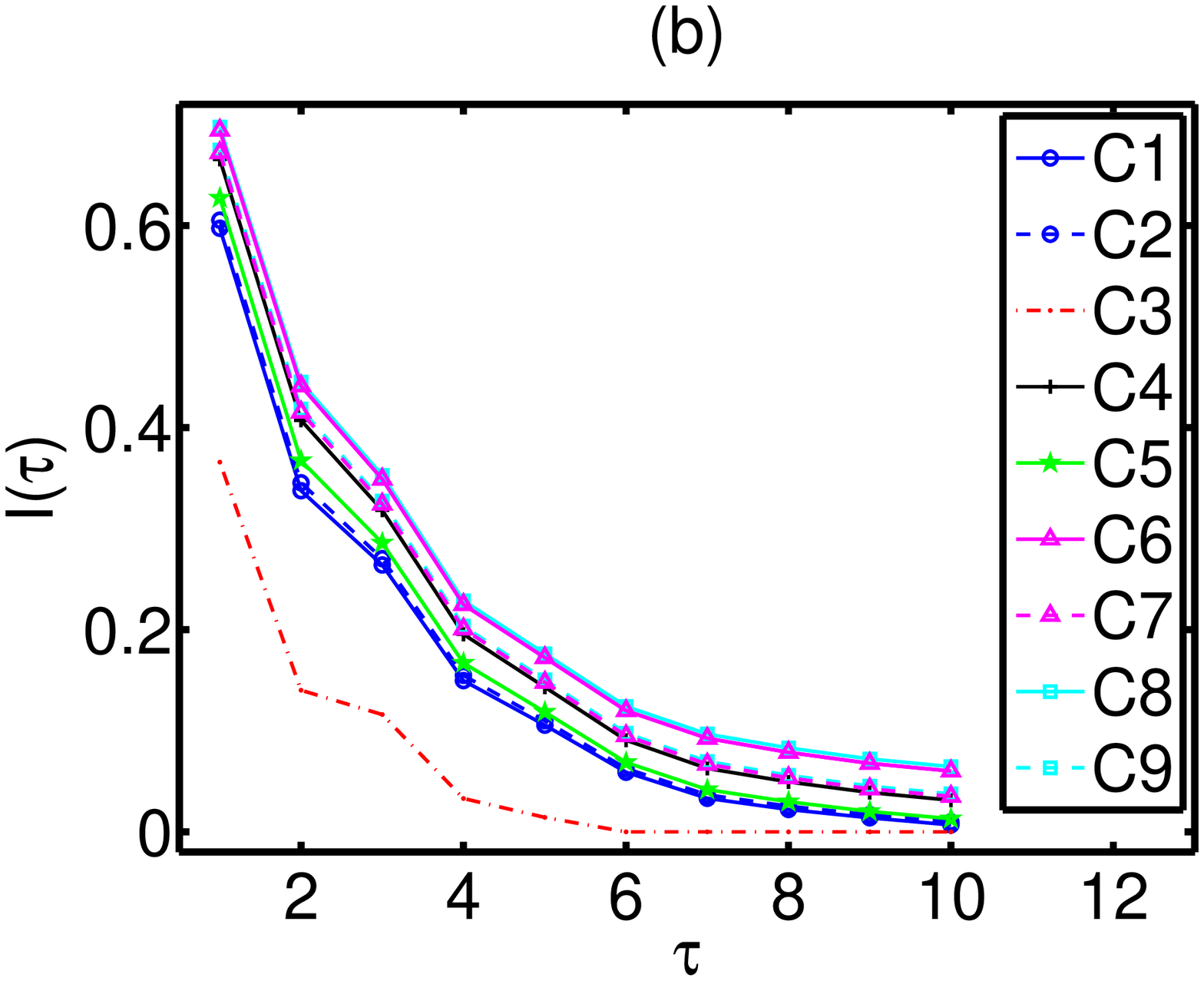}}}
\caption{A. Papana} \label{fig:HenonKEcrit}
\end{figure}

\clearpage
\begin{figure}
\centerline{\hbox{
\includegraphics[height=9cm,keepaspectratio]{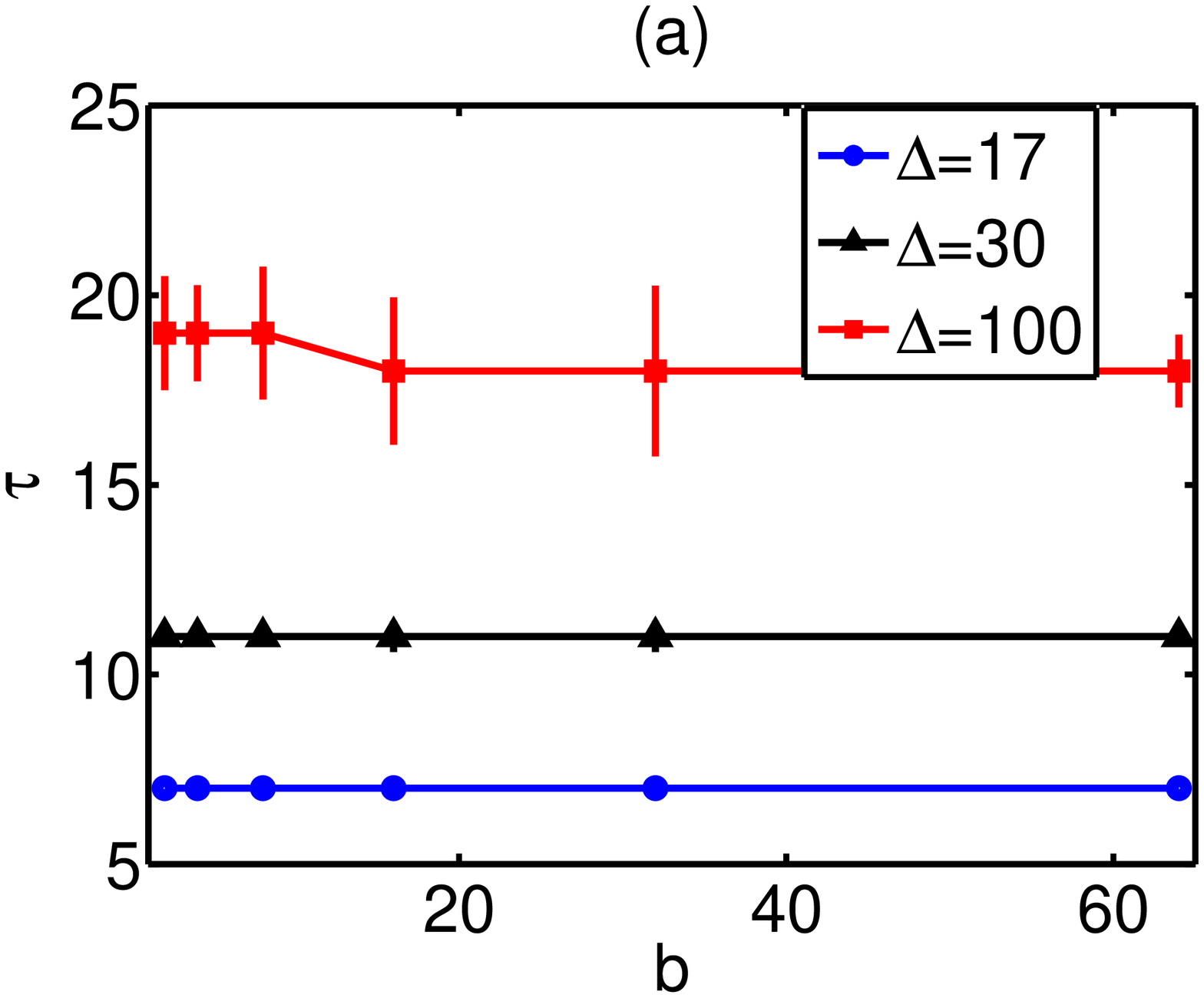}}}
\centerline{\hbox{
\includegraphics[height=9cm,keepaspectratio]{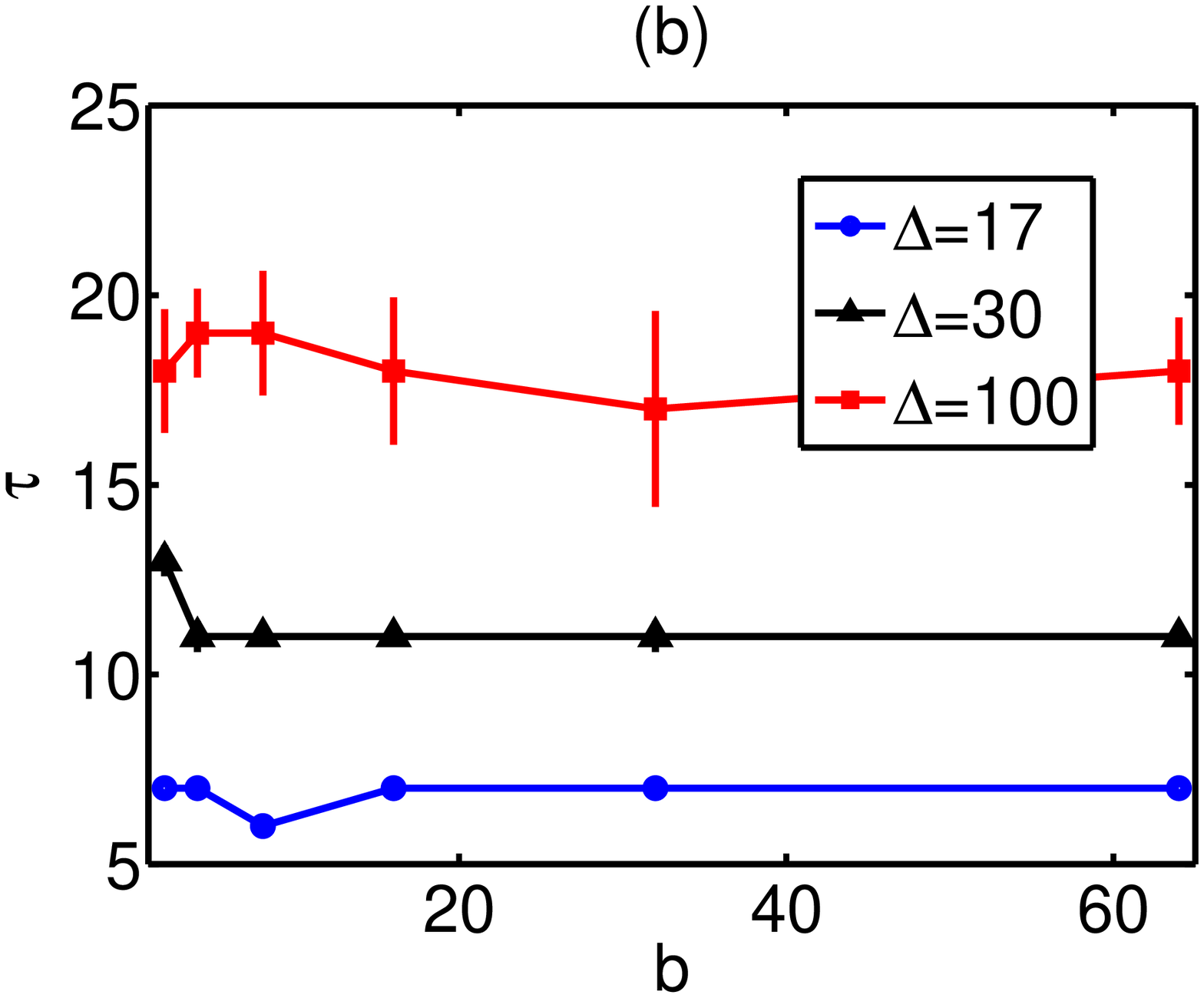}
}} \caption{A. Papana} \label{fig:MGlassminlag}
\end{figure}

\clearpage
\begin{figure}
\centerline{\hbox{
\includegraphics[height=9cm,keepaspectratio]{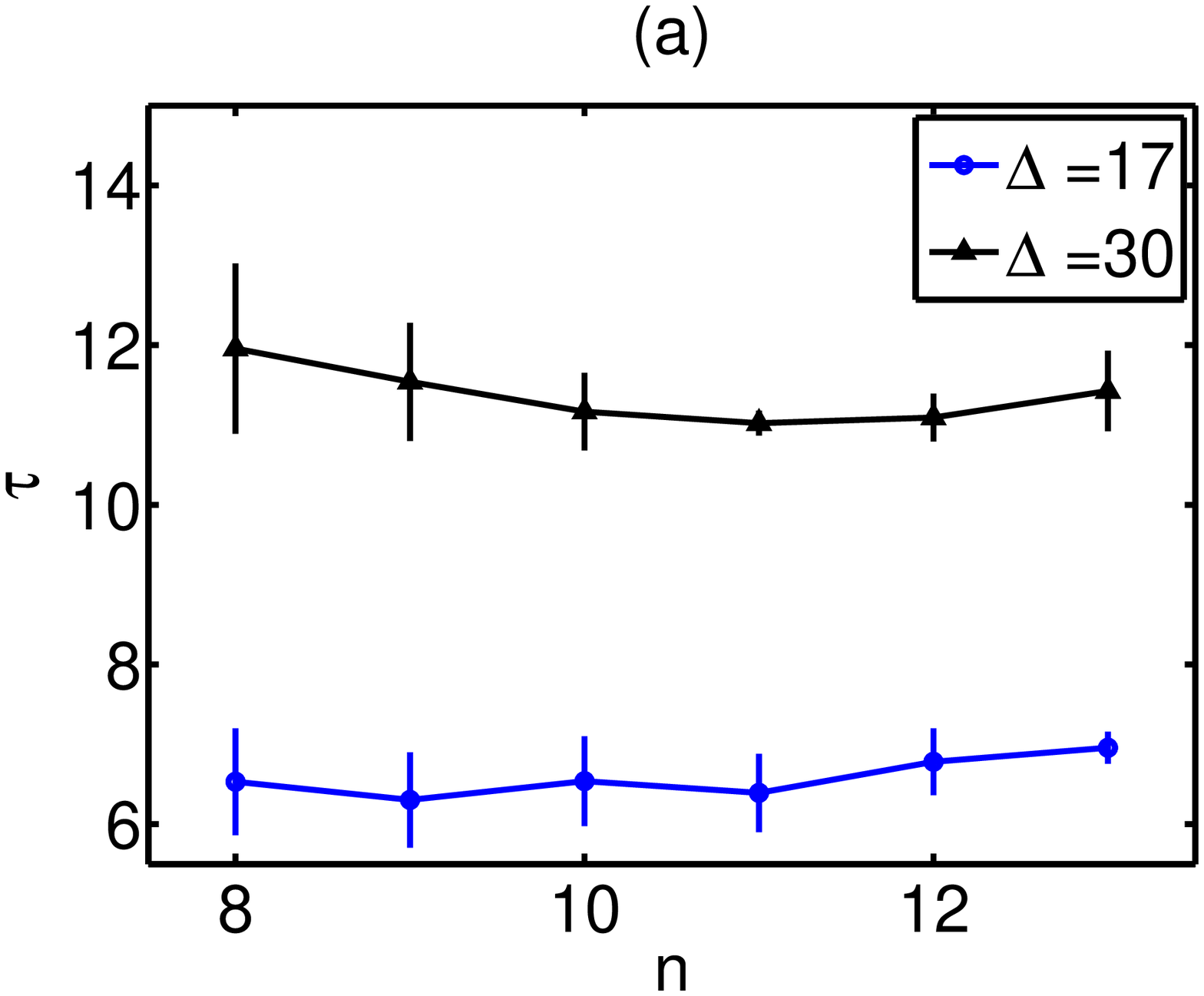}}}
\centerline{\hbox{
\includegraphics[height=9cm,keepaspectratio]{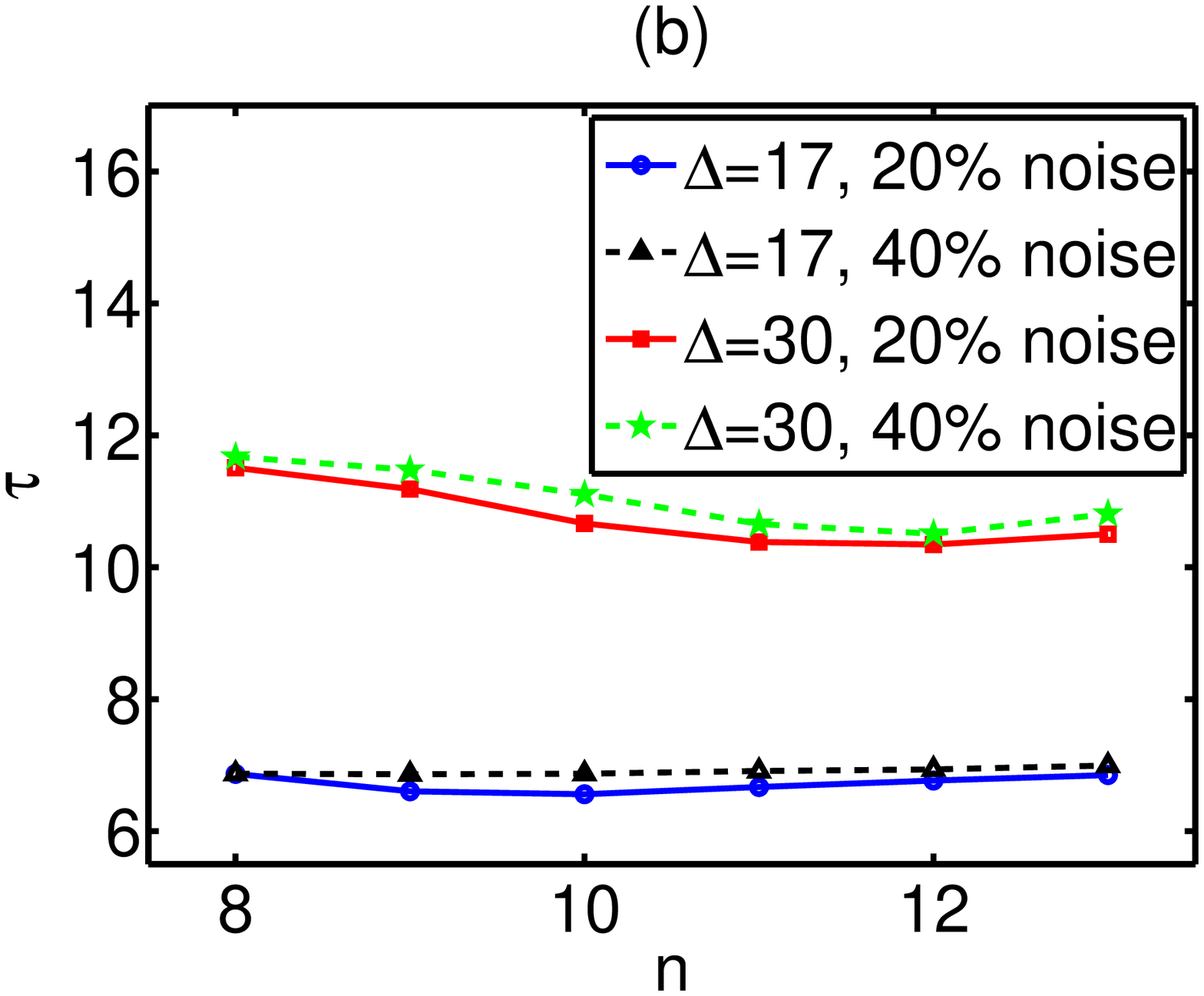}}}
 \caption{A. Papana}
\label{fig:ADMGLass}
\end{figure}

\clearpage

\begin{figure*}
\centerline{\hbox{
\includegraphics[height=9cm,keepaspectratio]{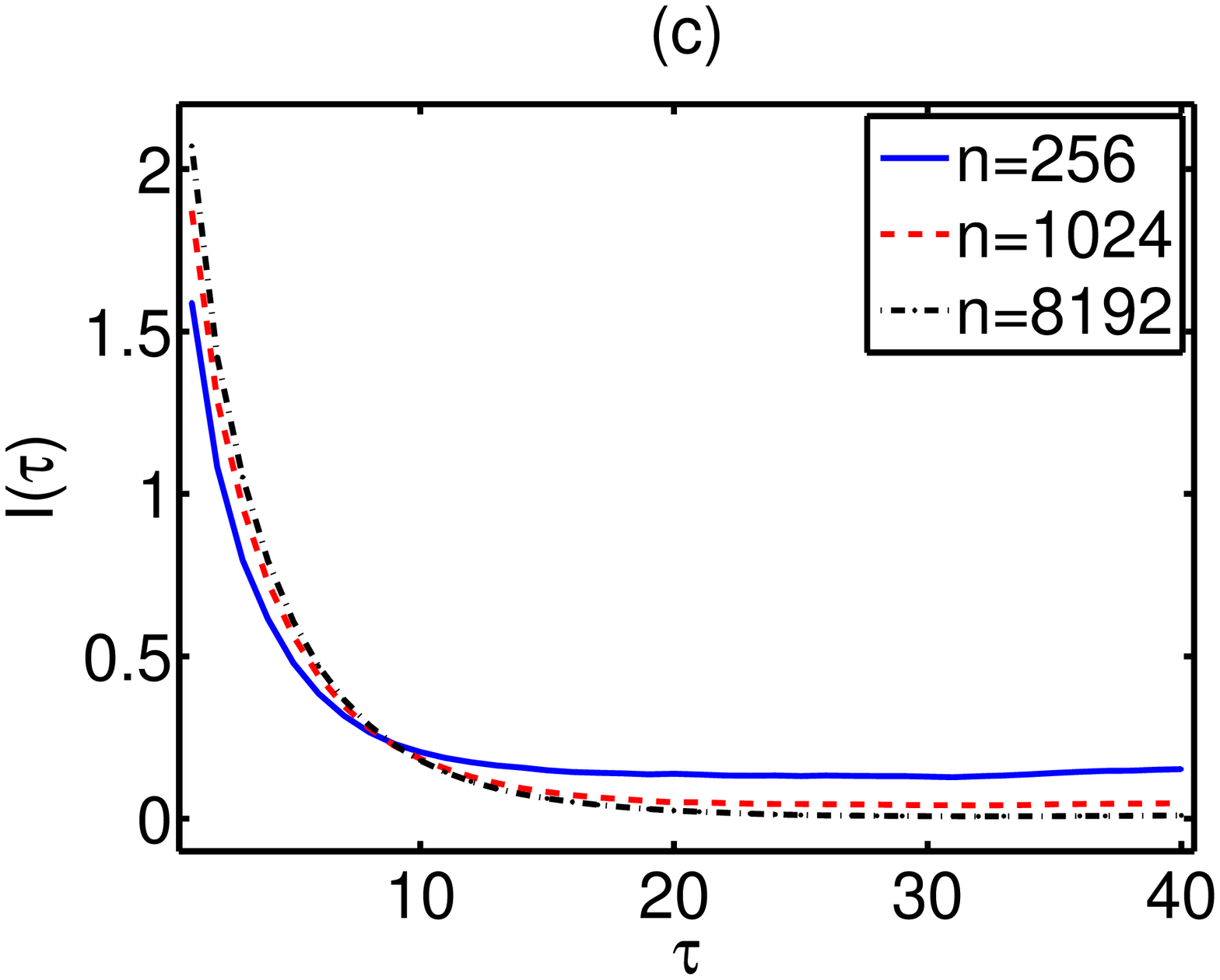}}}
\centering{Figure 13c: A. Papana}
\end{figure*}

\clearpage
\begin{figure}
\centerline{\hbox{
\includegraphics[height=9cm,keepaspectratio]{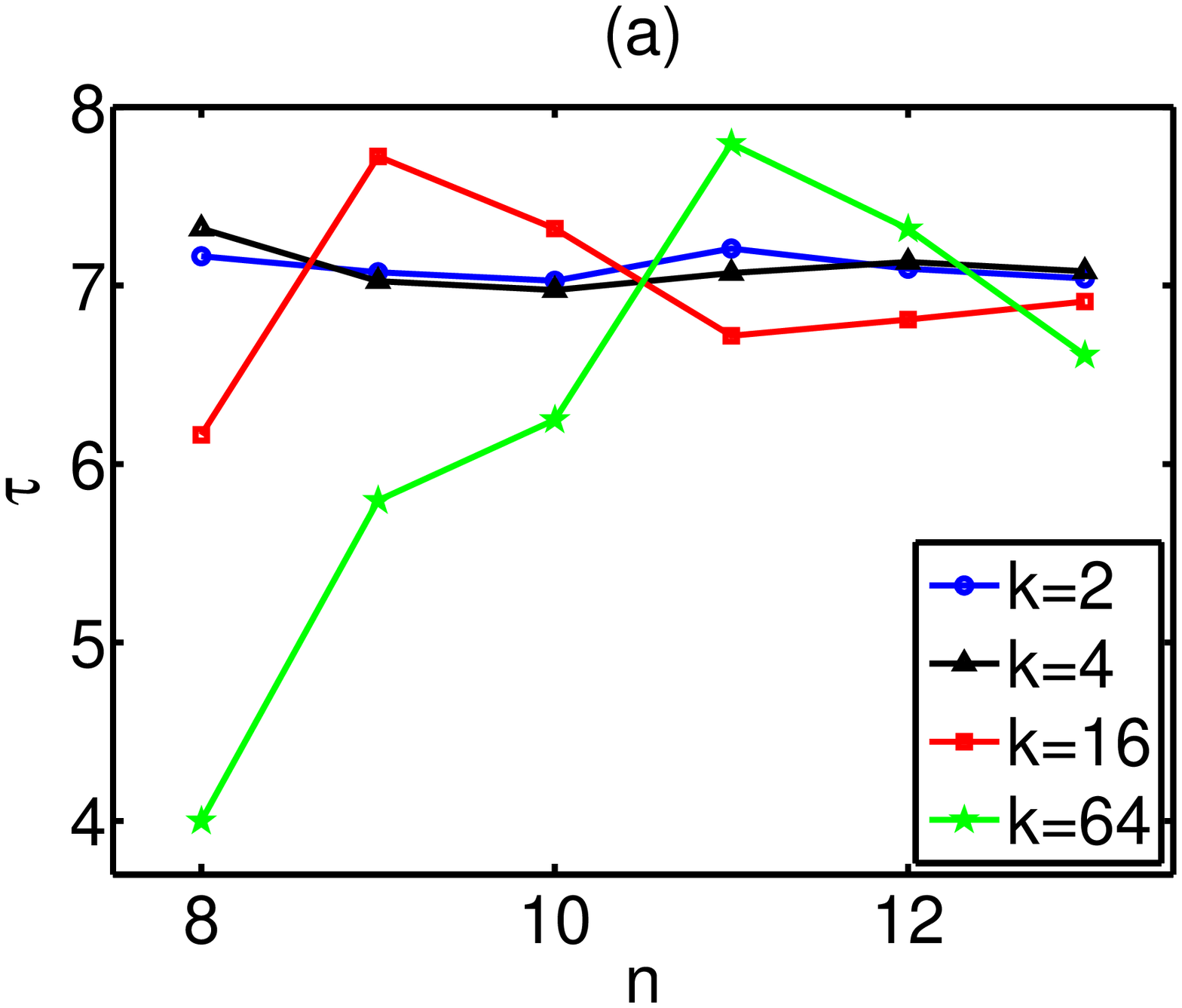}}}
\centerline{\hbox{
\includegraphics[height=9cm,keepaspectratio]{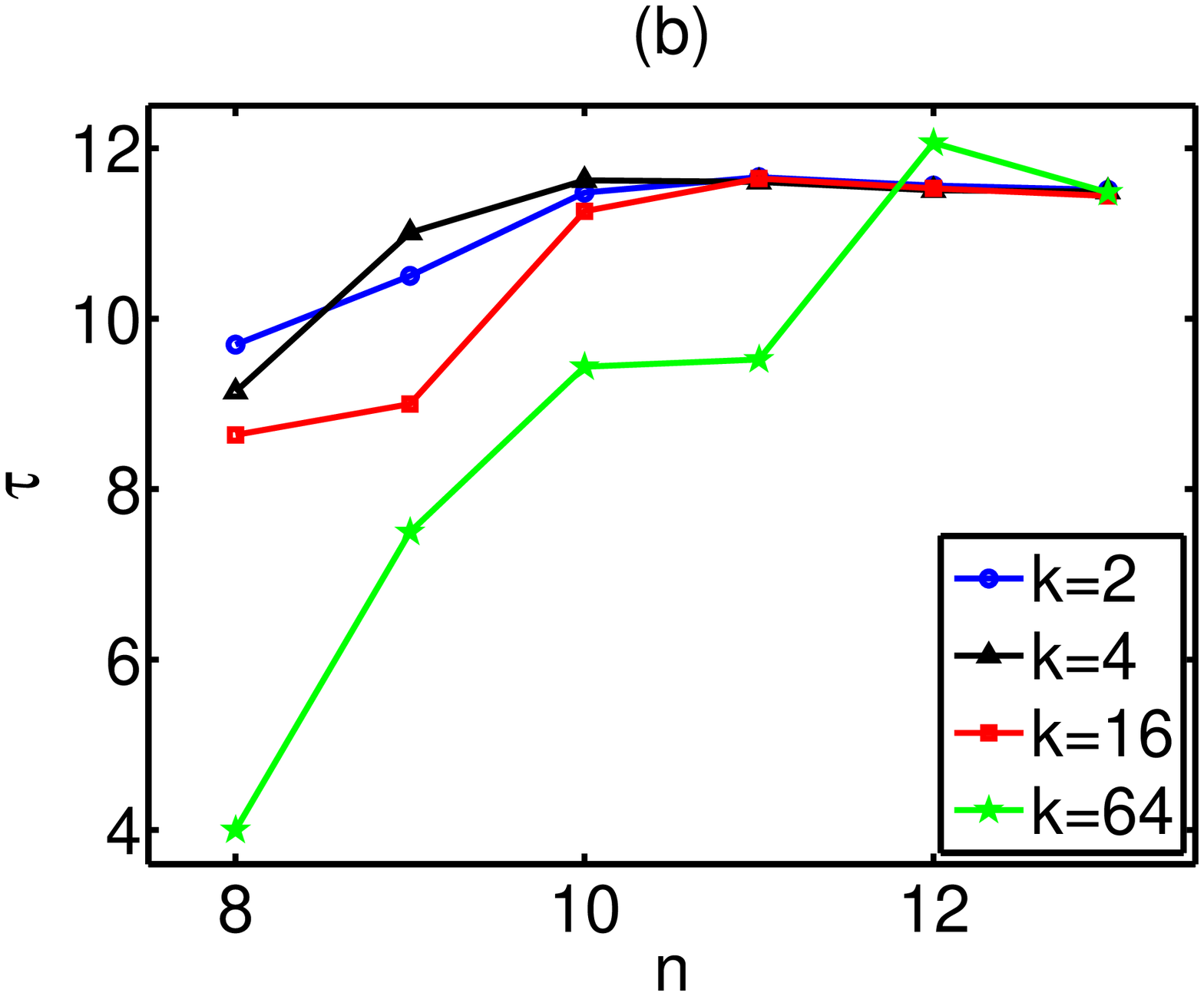}}}
\caption{A. Papana} \label{fig:MGlassKNN}
\end{figure}

\clearpage

\begin{figure*}
\centerline{\hbox{
\includegraphics[height=9cm,keepaspectratio]{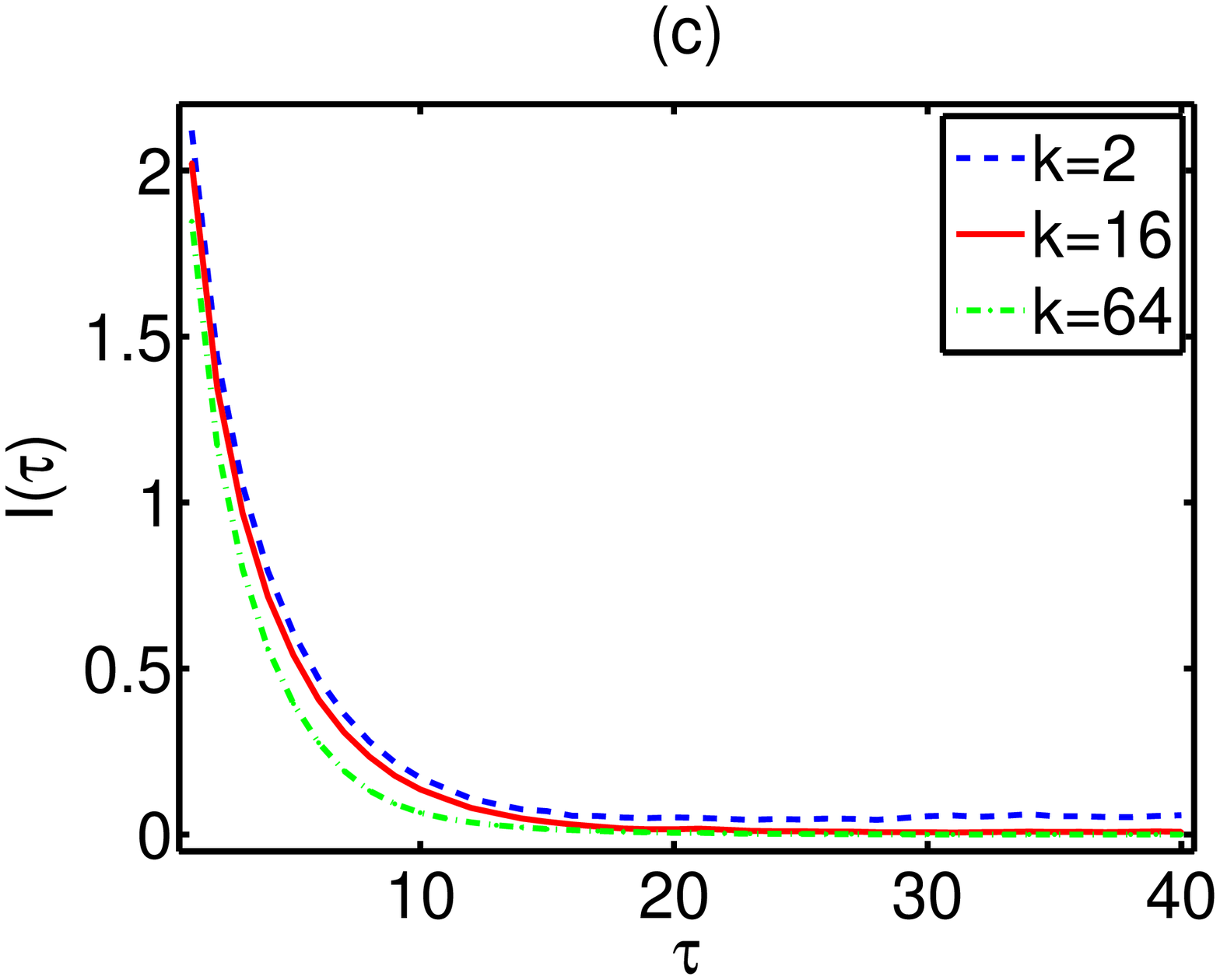}}}
\centering{Figure 14c: A. Papana}
\end{figure*}
\end{document}